\documentclass[jfm]{jfm}
\usepackage[applemac]{inputenc}
\usepackage{zefonts}           
\usepackage{amsmath}

\usepackage{color}
\usepackage{soul}

\ifCUPmtlplainloaded \else
  \checkfont{eurm10}
  \iffontfound
    \IfFileExists{upmath.sty}
      {\typeout{^^JFound AMS Euler Roman fonts on the system,
                   using the 'upmath' package.^^J}%
       \usepackage{upmath}}
      {\typeout{^^JFound AMS Euler Roman fonts on the system, but you
                   dont seem to have the}%
       \typeout{'upmath' package installed. JFM.cls can take advantage
                 of these fonts,^^Jif you use 'upmath' package.^^J}%
      }
  \else
  \fi
\fi

\ifCUPmtlplainloaded \else
  \checkfont{msam10}
  \iffontfound
    \IfFileExists{amssymb.sty}
      {\typeout{^^JFound AMS Symbol fonts on the system, using the
                'amssymb' package.^^J}%
       \usepackage{amssymb}%

      }{}
  \fi
\fi

\ifCUPmtlplainloaded \else
  \IfFileExists{amsbsy.sty}
    {\typeout{^^JFound the 'amsbsy' package on the system, using it.^^J}%
     \usepackage{amsbsy}}
    {}
\fi

\newsavebox{\astrutbox}
\sbox{\astrutbox}{\rule[-5pt]{0pt}{20pt}}

\ifx\pdfoutput\undefined      
\usepackage{graphicx}
\graphicspath{{figuresEPS/}}
\else                         
\usepackage[pdftex]{graphicx}
\usepackage{epstopdf}         
\graphicspath{{figuresEPS/}}     
\fi
\title{ {Pendulums}, Drops and Rods: a physical analogy}

\author[Benoît Roman, Cyprien Gay  and Christophe Clanet]
{Benoît Roman$^{1}$, Cyprien Gay$^{2}$ and Christophe Clanet$^{3}$}

\affiliation{$^1$PMMH, UMR7636 du CNRS, ESPCI, 10 rue Vauquelin, 75005 Paris, FRANCE.\\
$^2$MSC, UMR 7057 du CNRS, Université de Paris Diderot, France.\\
$^3$LadHyX, UMR7646 du CNRS, Ecole Polytechnique, 91128 Palaiseau, FRANCE.}

\begin{document}

\maketitle

\newcommand{\be}{\begin{equation}}
\newcommand{\ee}{\end{equation}}
\newcommand{\bee}{\begin{eqnarray}}
\newcommand{\eee}{\end{eqnarray}}
\newcommand{\hs}{\hspace{1cm}}
\newcommand{\LE}{L_{\rm el}}
\newcommand{\LC}{L_{\rm cap}}

\textbf{\large Abstract.}\\

 A liquid meniscus, a bending rod (also called elastica)
and a simple pendulum are all
 described by the same non-dimensional equation.
 The oscillatory regime of the pendulum corresponds to buckling
 rods and pendant drops,
 and the high-velocity regime corresponds to
 spherical drops, puddles and multiple rod loopings.
 We study this analogy in a didactic way and discuss  {how,} despite this
 common governing equation, the three systems are not completely
 equivalent. We also consider the cylindrical deformations
of an inextensible, flexible membrane containing a liquid,
which in some sense interpolates
between the meniscus and rod conformations.

 \section{Introduction}
 {C}hapter LXXXIII of the Encyclopoedia 
Britannica (ninth edition) was written by JC Maxwell probably in the
eighteenseventies, as he was organizing the the new Cavendish Laboratory\footnote{JC Maxwell 
died in 1879 at the age of 48 years.}. The whole article is 50 pages 
long, all dedicated to Capillary Action. Organized in 18 subchapters, 
the 7th is dedicated to the shape of a liquid meniscus on 
a wall:

\noindent
\emph{The form of this capillary surface is identical with that of 
the ``elastic curve'', or the curve formed by a uniform spring 
originally straight, when its ends are acted on by equal and opposite 
forces applied either to the ends themselves or to solid pieces 
attached to them. Drawings of the different forms of the curve may be 
found in Thomson and Tait's Natural Philosophy, Vol.I, p 455.}

\noindent
This sentence has motivated the present work. 

\noindent
The power of the analogies in Physics is to enlight one subject 
 using the known properties of its analogue. Here, 
 pendulums, drops and rods are all classical subjects and the purpose 
 is not to enlighten any one of the topics but to emphasize the properties 
 shared by these three different systems. 

 \section{The pendulum}
\begin{figure}
    \centering
    \includegraphics*[scale=0.7]{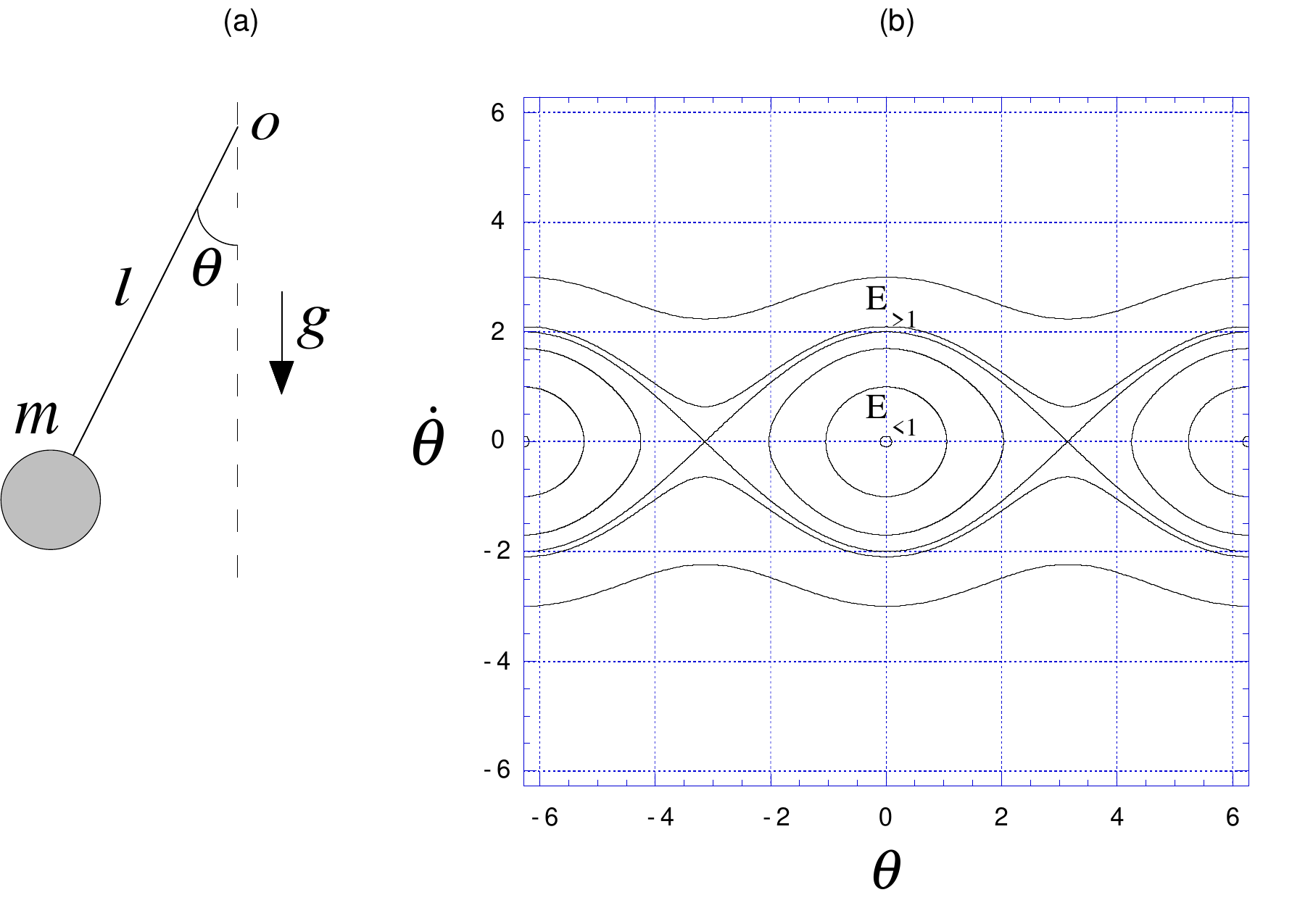}
    \caption{Pendulum:
    (a)  {sketch} of the experiment,
    (b) Trajectories of the pendulum
    in the $[\theta, \dot{\theta}]$ phase diagram.}
   \label{Figure01}
\end{figure}

A pendulum is presented on figure~\ref{Figure01}-(a).
A particle of mass $m$ moves under the gravitational acceleration $g$,
remaining at a constant distance $l$ from the fixed point $o$.
The characteristic time of the system is:
\begin{equation}
    \tau=\sqrt{l/g}.
    \label{Eq1}
\end{equation}
Denoting by $\theta$ the angle between the vertical and the pendulum,
the non-dimensional equation of motion reads:
\begin{equation}
    \frac{d^2\theta}{dt^2}=-\sin\theta.
    \label{eqp1}
\end{equation}
The minus sign 
is changed to a plus sign when gravity is reversed.
Equation (\ref{eqp1}) is then recovered in its original form
through the translation $\theta\rightarrow\theta+\pi$.
Equation (\ref{eqp1}) is thus generic for the pendulum,
whatever the direction of the gravity field.
The first integral of motion of this equation corresponds to the
conservation of the total energy
(in units of $mgl$):
\begin{equation}\label{enerp}
E\equiv\frac12\dot{\theta}^2-\cos\theta
\end{equation}
The trajectories obtained with different energy values are
presented in the phase diagram $[\theta,\dot{\theta}]$ on figure
\ref{Figure01}-(b). Two different regions appear: a region with
closed orbits, denoted by $E_{<1}$, where the total energy is
smaller than unity, and a region with open trajectories, $E_{>1}$,
where the total energy is larger than unity.  The main difference
between both regions is the existence  on the trajectories of
points with vanishing angular speed $\dot{\theta}$ in the region
$E_{<1}$, and their absence in the region $E_{>1}$. We will see
later that these points correspond to inflexion points on the
shape of a rod, or of a drop. The special case $E=1$ (separatrix)
defines the soliton motion and will be addressed separately in
section~\ref{MLS}.

  \section{Drops and bubbles}
\begin{figure}
    \centering
    \includegraphics*[scale=0.7]{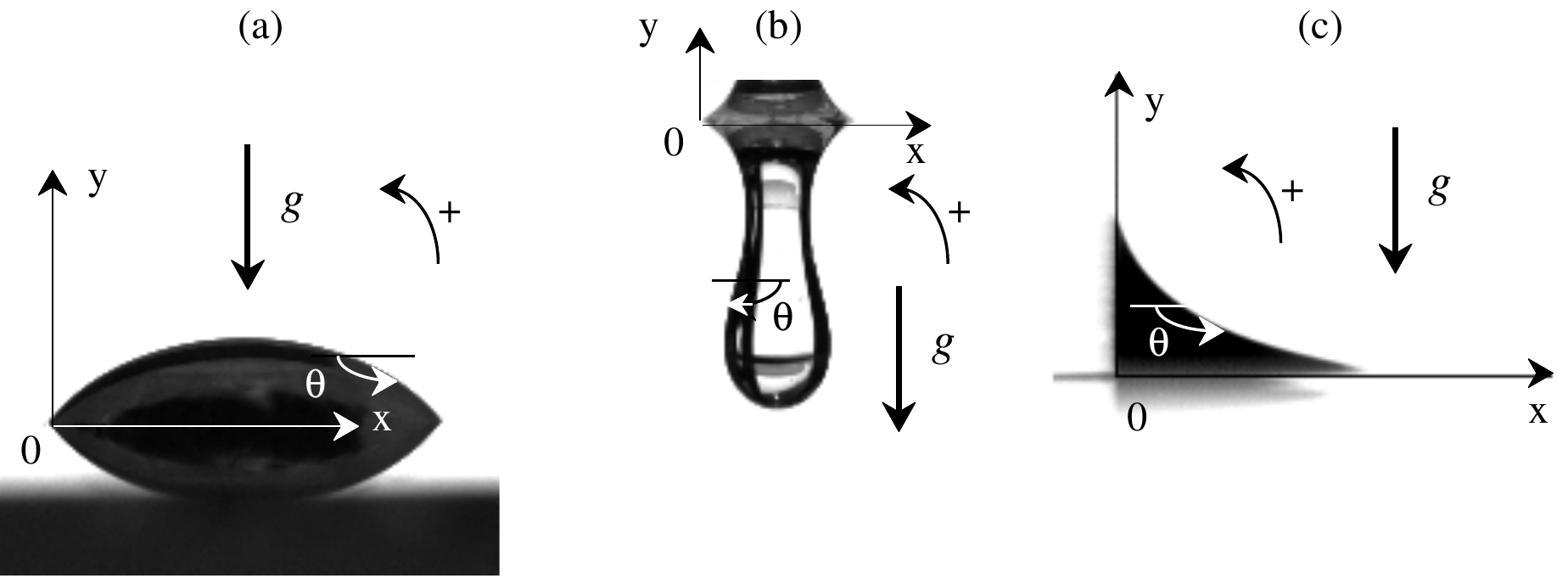}
    \caption{2D drops and meniscus:
    (a) drop compressed by gravity and its image reflected on the
substrate,
    (b) pending drop streched by gravity,  {(note the different convention for measurement of angle $\theta$ in this case, which allows to keep the same equations throughout the article)}
    (c) meniscus near a solid wall. }
   \label{Figure02}
\end{figure}

In this section, we consider the shapes of 2D static 
drops\footnote{2D is used to designate shapes with one curvature. Axisymmetric 
shapes are thus not included  {and will be discussed in section \ref{3Ddrops}. } }, either
compressed (figure~\ref{Figure02}-(a)), or streched by gravity
(figure~\ref{Figure02}-(b)). The 2D meniscus presented in figure
\ref{Figure02}-(c) will be discussed in section~\ref{MLS}. In these
three cases, the shape of the interface results from the
equilibrium between surface tension, $\sigma$, and gravity, $g$.
We describe this equilibrium with
the notations and conventions presented in figure~\ref{Figure02}.
Note that the definition adopted for $\theta$
is different for drops compressed by gravity (and the
special case of meniscus) and for drops stretched by gravity.
With these conventions, we will deal with the same equation
throughout our discussion. In the three cases, $\theta$ is measured 
through the liquid from the horizontal to the tangent.

The equilibrium shape of a bubble blocked beneath a solid surface
(resp. anchored to a solid)
can be mapped, through reflection by a horizontal plane,
onto that of a liquid drop resting on a plane (resp. a pendant drop).
Indeed, gravity and density difference are both
reversed through this symmetry.
In this section, we will therefore simply focus on drops.
\subsection{Drops on a solid surface.}

\begin{figure}
    \centering
    \includegraphics*[scale=0.7]{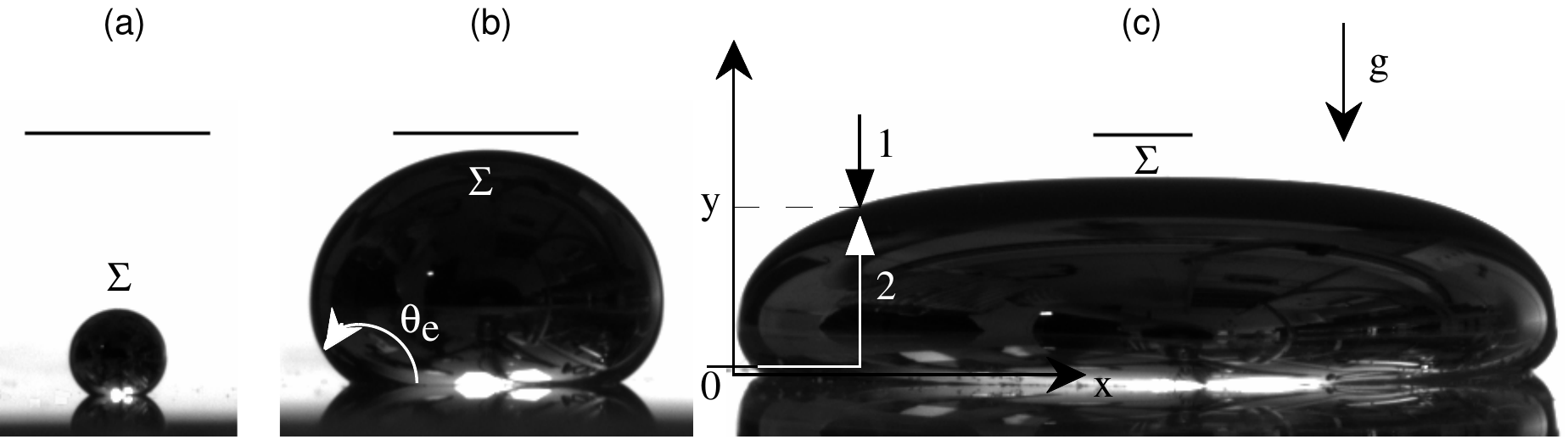}
    \caption{Drops of mercury on plexiglass ($\theta_{e}\approx\pi$),
    with different volumes:
    (a) $\mathcal{V} = 0.35\mbox{ mm$^{3}$}$, 
    (b) $\mathcal{V} = 12.9\mbox{ mm$^{3}$}$, with  {$\theta_e$ is the contact angle, (defined with standard angle convention)}
    (c) $\mathcal{V} = 426.5\mbox{ mm$^{3}$}$,
    (the horizontal bar on each picture represents 1.7 mm). }
   \label{Figure03}
\end{figure}

The shape of a liquid drop resting on a solid surface
is characterized by its volume, $\mathcal{V}$, and the contact angle,
$\theta_{e}\in [0,\pi]$ between the free surface and the solid.
Small contact angles correspond to good wetting
(if the contact angle is zero, wetting is said to be complete),
while large contact angles are observed for poor wetting
or non-wetting situations.
Three drops of mercury on plexiglass ($\theta_{e}\approx\pi$),
are presented in figure~\ref{Figure03}. As the volume is increased, the
shape changes from an almost perfect circle
(figure~\ref{Figure03}-(a)) to a puddle (figure~\ref{Figure03}-(c)).

\begin{figure}
     \centering
      \includegraphics*[scale=0.6]{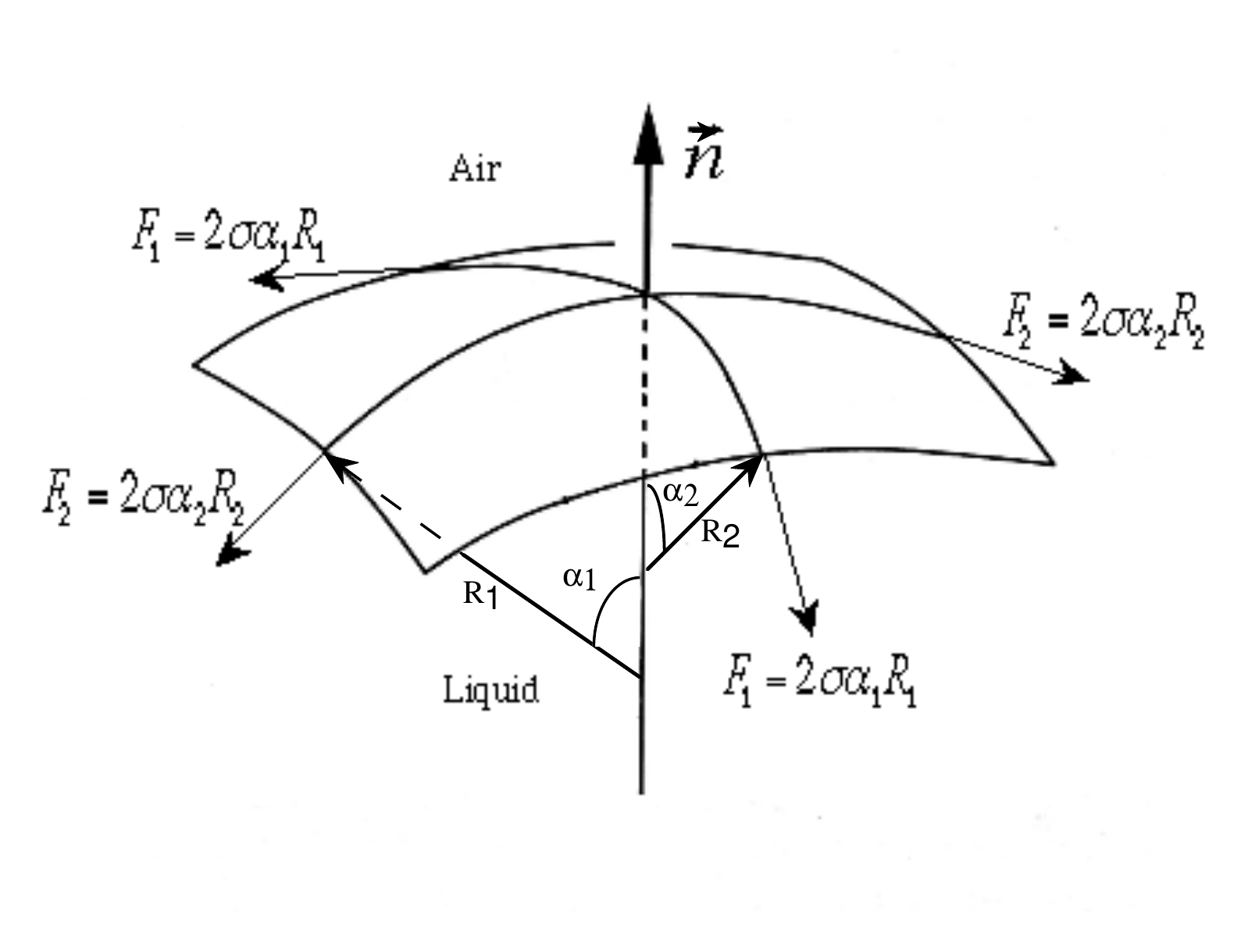}
    \caption{Illustration of the Laplace pressure jump $\Delta p_L$
        across a curved interface.
 We consider a small element of surface
 whose principal radii of curvature are $R_1$ and $R_2$
 and which spans an infinitesimal angle $2\alpha_1$ in one direction
 and $2\alpha_2$ in the other direction,
 as seen from the centers of curvature.
 Let $F_1=2\sigma\alpha_1R_1$ is the force
 due to the surface tension $\sigma$
 acting on one edge of length $2\alpha_1R_1$.
 It is oriented at angle $\alpha_2$ from the tangent plane to the surface.
 Its normal component is thus $F_1\sin\alpha_2\approx F_1\alpha_2$.
 Including the opposite edge and the similar contribution from $F_2$,
 we obtain $\Delta
p_L.2R_1\alpha_1.2R_2\alpha_2=4\alpha_1\alpha_2\sigma(R_1+R_2)$,
 which justifies the expression of the Laplace pressure jump $\Delta
 p_L$ given by equation (\ref{laplace}).}
    \label{Figure04}
 \end{figure}

To account for the shape, we express the pressure
at a given point located behind the interface at altitude $y$
in two ways, following the paths~(1) and~(2)
indicated on figure~\ref{Figure03}-(c).
The pressure, $p$, beneath the interface is larger than the
ambient pressure, $p_{0}$, due to surface tension,
as can be seen from figure~\ref{Figure04}.
When the free surface is curved, the force associated to surface tension 
is tangent to the surface and changes directions from place to place. As a result,
the force is not balanced on both sides of a curved section of the
surface. Hence, it exerts a pressure towards the center of
curvature ($-\overrightarrow{n}$ direction),
which results in a pressure difference between the liquid and the
gas, called the Laplace pressure, $\Delta p_{L}$, which is
proportional to the total curvature of the surface:
\begin{equation}
    \Delta p_{L}=\sigma\left(\frac{1}{R_{1}}+\frac{1}{R_{2}}\right).
    \label{laplace}
\end{equation}
where $R_1$ and $R_2$ are the principal radii of curvature. 
In figure~\ref{Figure04}, $R_{1}$ and $R_{2}$ are both positive and 
surface tension imposes that the 
pressure in the liquid is higher than in the air.
In a two-dimensional situation where the shape
is invariant in one direction,
one principal radius of curvature is infinite,
and the total curvature is simply  {equal to} the curvature
of the curve in the perpendicular cross-section.
Laplace's equation thus yields the pressure $p$
following path~(1):
\begin{equation}
    p=p_0-\sigma \frac{d\theta}{ds},
    \label{laplace00}
\end{equation}
where $p$ denotes the pressure in the liquid, and $s$ is a  
curvilinear coordinate. In this equation, $d\theta/ds$ is the local 
curvature of the interface.
Following path~(2), the pressure contains also a hydrostatic term:
\begin{equation}
    p=p_0-\sigma \left(\frac{d\theta}{ds}\right)_{0}-\rho g y,
    \label{hydrostatics}
\end{equation}
where $\rho$ is the density of the liquid,
$y$ is the altitude  {and $\left(\frac{d\theta}{ds}\right)_{0}$ is the curvature at height $y=0$}. From
equations~(\ref{laplace00}) and~(\ref{hydrostatics}):
\begin{equation}
    \frac{d\theta}{ds}=\left(\frac{d\theta}{ds}\right)_{0}+\frac{y}{a^2},
    \label{equality}
\end{equation}
where the characteristic length scale
\begin{equation}\label{aliq}
a\equiv\sqrt{\sigma/(\rho g)}
\end{equation}
is the capillary length.
Non-dimensionalization by $a$ and differentiation
with respect to $s$
yields the pendulum equation~(\ref{eqp1}):
\begin{equation}
    \frac{d^2\theta}{ds^2}=-\sin\theta,
    \label{drop-pendulum1}
\end{equation}
where the geometrical relation $dy/ds=-\sin\theta$ has been used.
 {As compared to equation~(\ref{eqp1}),} 
the time derivative of the pendulum equation has been
replaced by the spatial derivative along the surface,
and the characteristic time $\tau=\sqrt{l/g}$
has been replaced by the characteristic length $a=\sqrt{\sigma/(\rho g)}$.
Since both phenomena share the same non-dimensional equation,
their phase diagrams are also identical.
The total energy,
\begin{equation}\label{enerm}
E\equiv\frac12\dot{\theta}^2-\cos\theta
\end{equation}
(which is the same as equation~(\ref{enerp}) for the pendulum),
can be evaluated at the apex, $\Sigma$, indicated on figure \ref{Figure03}.
By symmetry, $\theta_{\Sigma}=\pi$, so that $E=1+\dot{\theta}_{\Sigma}^2/2$.
For all the drops lying on a solid surface, the energy $E$ is thus larger
than unity so that the corresponding trajectories always
lie in the $E_{>1}$ region,
where no inflection point are present
(the curvature, $\dot{\theta}$, does not vanish).

\begin{figure}
    \centering
    \includegraphics*[scale=0.7]{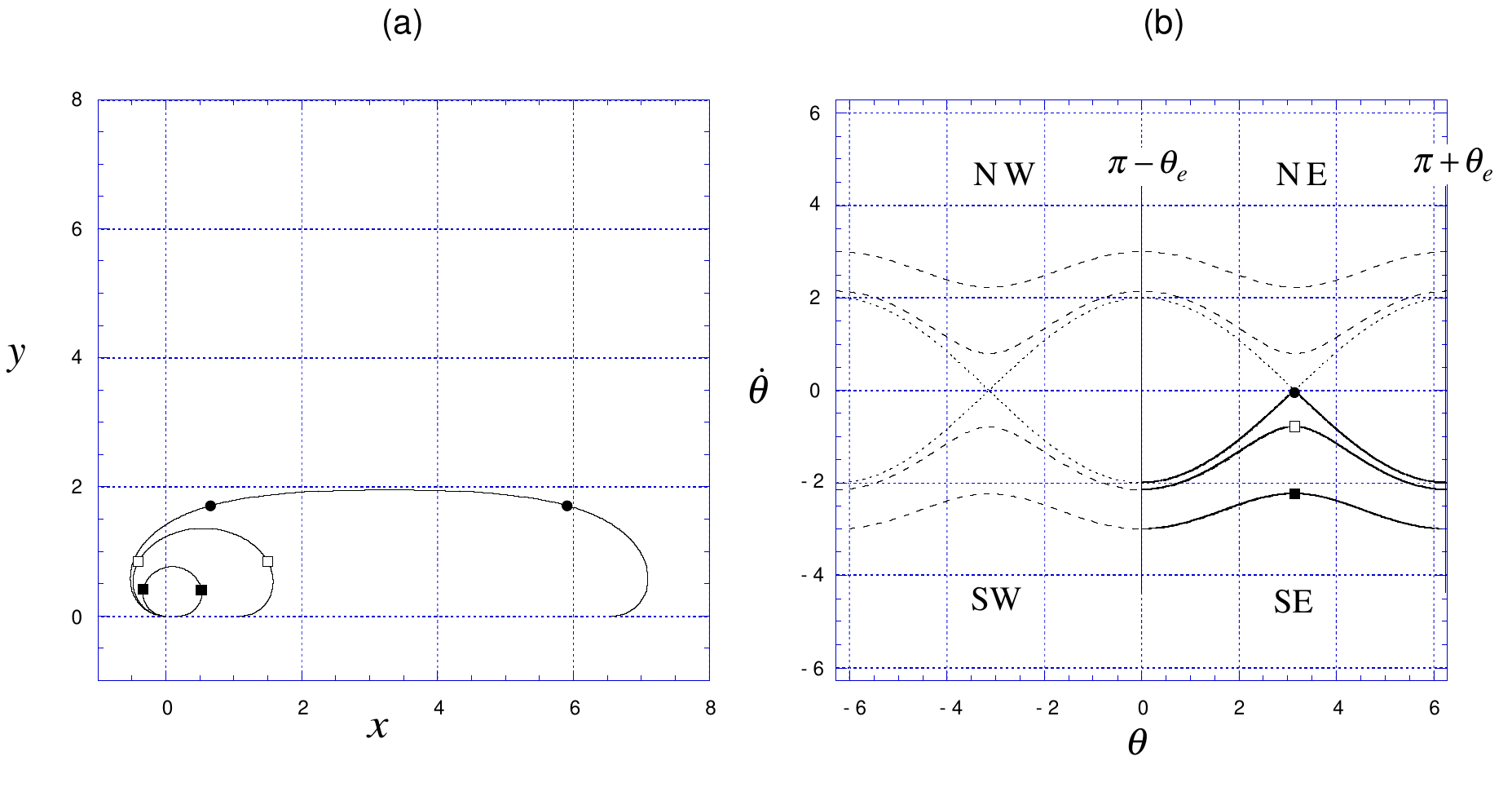}
    \caption{Numerical integration of equation (\ref{drop-pendulum1}),
    with $\theta(0)=2\pi$ and different energies,
    $\blacksquare$ $E=3.5$, $\square$ $E=1.3$, $\bullet$ $E=1.007$:
    (a) shape of the drops,
    (b) phase diagram.
Note that the $\bullet$ curve does not correspond exactly
to the separatrix ($E=1$) for which the drop
would have an infinite lateral extension (infinite puddle).}
    \label{Figure05}
\end{figure}
Figure~\ref{Figure05}-(a) shows the shapes obtained through the numerical
integration of equation (\ref{drop-pendulum1}), with $\theta(0)=2\pi$
and different energies, $E$, chosen so as to give the same aspect ratio as the
drops presented in figure~\ref{Figure03}. The corresponding trajectories
in the phase diagram $[\theta,\dot{\theta}]$ are presented in
figure~\ref{Figure05}-(b). The puddle approaches the separatrix defined
by $E=1$, whereas the small circular drop corresponds to the high energy
region. The limit of circular drops corresponds to the high energy
pendulum, where gravity almost does not affect the momentum of the pendulum.
With the conventions adopted on figure~\ref{Figure02}-(a),
the trajectory of one drop extends from $\pi-\theta_{e}$ to $\pi + \theta_{e}$
 (that is from $0$ to $2\pi$ with mercury on
plexiglass) and $\dot{\theta}$ is negative, so that all the drops are
contained in the south-east region of the phase diagram indicated with
symbols in figure \ref{Figure05}-(b). It is easy to show that with
different conventions keeping the equation (\ref{drop-pendulum1})
unchanged, the other three regions (SW, NW and NE) can be reached, one
at a time. In all cases, the trajectory is
symmetrical with respect to the vertical line $\theta=\theta_\Sigma$,
where $\theta_\Sigma$ corresponds to the apex of the drop.
At this point $\theta_\Sigma$, the
curvature $|\dot{\theta}|$ is minimal, due to the direction of gravity.
It clearly follows that $\theta_\Sigma=\pi$ modulo $2\pi$.

\subsection{Drops under a solid surface and pending drops.}

\begin{figure}
    \centering
    \includegraphics*[scale=0.7]{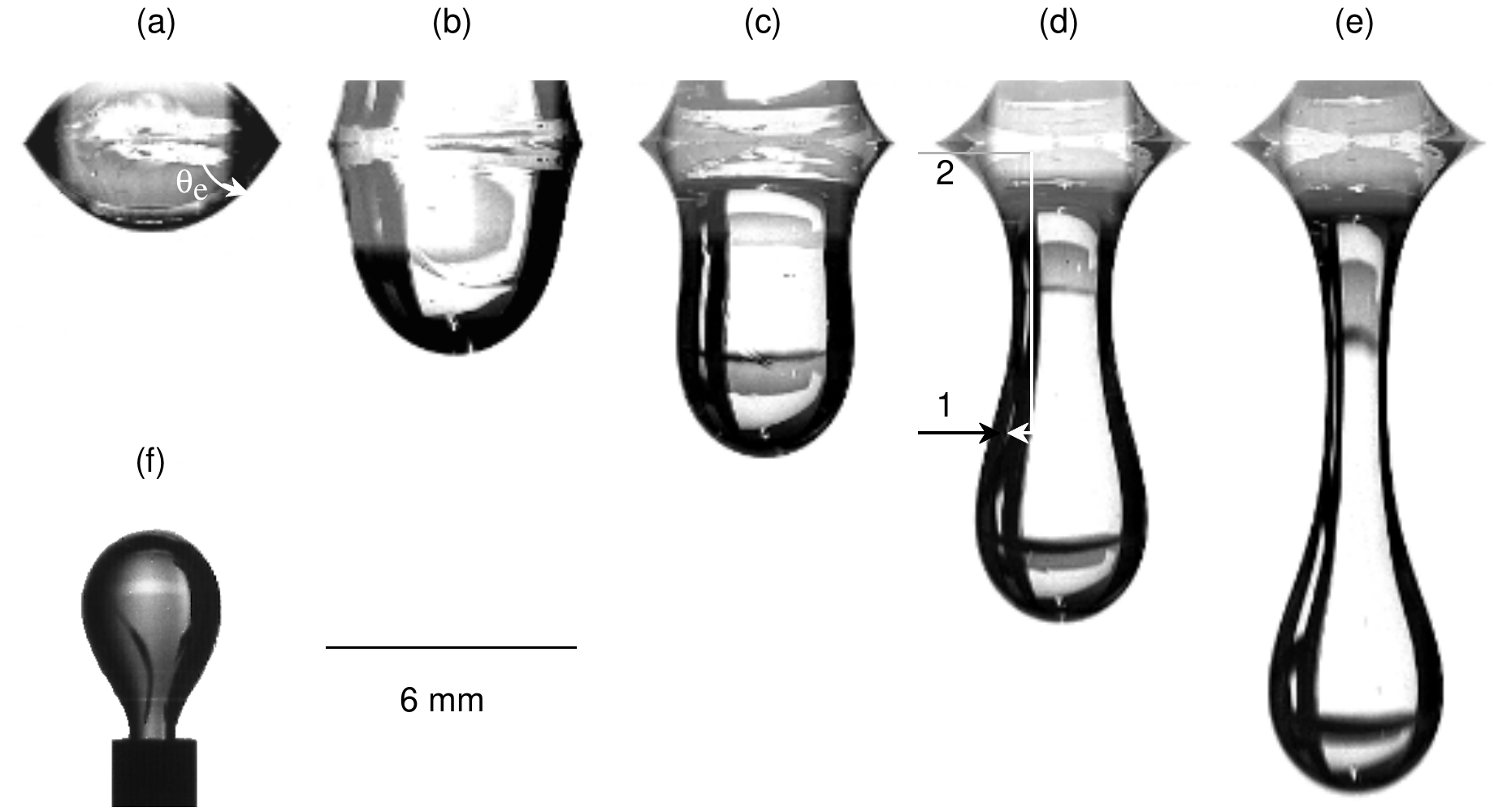}
    \caption{Drops and bubble.
Drops of glycerine under plexiglass with different volumes:
    (a) $\mathcal{V}=31\mbox{ mm$^3$}$  {(contact angle $\theta_e$ presented with usual angle definition)} ,
    (b) $\mathcal{V}=82\mbox{ mm$^3$}$, 
    (c) $\mathcal{V}=95\mbox{ mm$^3$}$,
    (d) $\mathcal{V}=95\mbox{ mm$^3$}$,
    (e) $\mathcal{V}=95\mbox{ mm$^3$}$.
(f) Anchored air bubble stretched by buoyancy in water.
The last three drops (c-e) are actually slowly deforming over time.
Their shapes are therefore not defined by the static equation,
since viscous stresses are involved.}
    \label{Figure06}
\end{figure}

\begin{figure}
    \centering
    \includegraphics*[scale=0.7]{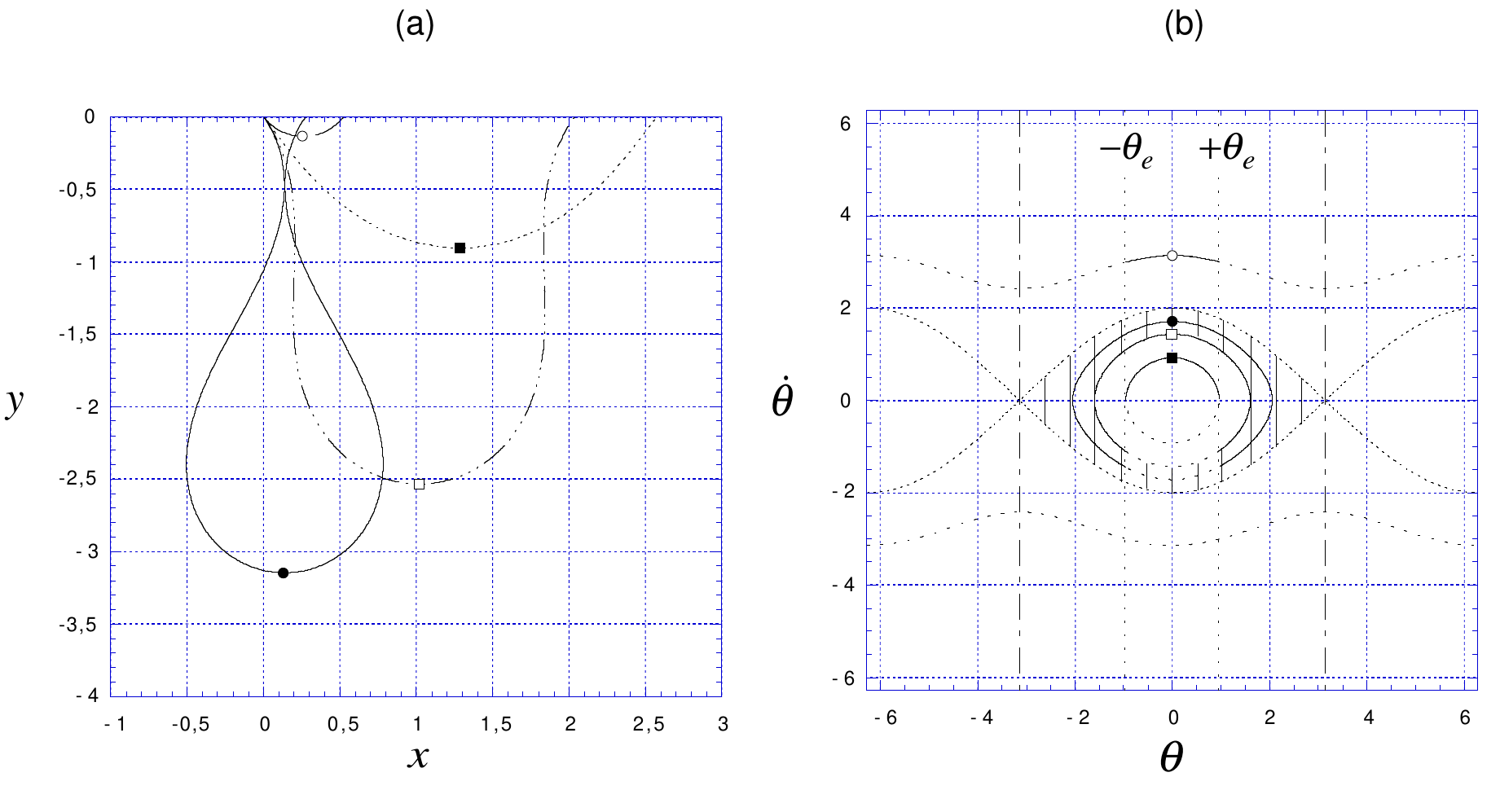}
    \caption{Shapes obtained from equation (\ref{drop-pendulum1})
    with $\theta(0)=-\theta_{e}=-55^{\circ}$ and different energy values,
    $\blacksquare$ $E=-0.57$,
    $\square$ $E=0.027$,
    $\bullet$ $E=0.468$,
    $\circ$ $E=3.92$:
     {(a) Shapes in real space
    (b) Corresponding trajectories in phase space. The dashed region is forbidden because 2D drops would
    include a larger volume than allowed by gravity.}}
    \label{Figure07}
\end{figure}

Figure~\ref{Figure06} shows drops of glycerine
hanging from a plexiglass surface
(contact angle $\theta_{e}\approx 55^{\circ}$).
The reflection on the plate allows for a
precise determination of both the angle of contact and the
position of the surface.
Such three-dimensional drops or bubbles are described
by a slightly more complex equation
(see~(\ref{exactdrop}) in the next paragraph)
than the 2D equation that we shall now derive.
With the conventions\footnote{ {Note that the definition of angle $\theta$ used in this section and in  figure~\ref{Figure02}-(b)  is related to that of figure~\ref{Figure02}-(a) by $\theta \to \pi+\theta$. This allows to use the same equation
for all systems in the article.} } of figure~\ref{Figure02}-(b),
the pressure at a given point beneath the interface
can be expressed in two ways
(paths~(1) and~(2) of figure~\ref{Figure06}-d). Through
path~(1), we get:
\begin{equation}
    p=p_0+\sigma \frac{d\theta}{ds},
    \label{laplace1}
\end{equation}
and through path~(2):
\begin{equation}
    p=p_0+\sigma \left(\frac{d\theta}{ds}\right)_{0}-\rho g y.
    \label{hydrostatique1}
\end{equation}
Upon non-dimensionalisation by the capillary length and
differentiation with respect to the curvilinear distance $s$,
we recover the pendulum equation (\ref{drop-pendulum1}),
since $dy/ds=\sin\theta$ in this case.
Figure~\ref{Figure07}-(a) shows
four shapes obtained from equation~(\ref{drop-pendulum1}) with
$\theta(0)=-\theta_{e}$ and different values of the energy.
The corresponding trajectories in the phase diagram $[\theta,\dot{\theta}]$
are presented in figure~\ref{Figure07}-(b). These
trajectories extend from $-\theta_{e}$ to $+\theta_{e}$,
and some of them have an inflection point.
Due to the orientation of gravity,
$|\dot{\theta}|$ is maximal on the axis of symmetry,
which implies that it is located at $\theta=0$ modulo $2\pi$.
Note that the anchored bubble (figure~\ref{Figure06}-(f))
 {is the symmetric}  {counterpart}  {of the pendant drop}.

Pendant drops cannot exceed some maximal volume, $\mathcal{V}_{max}$,
beyond which capillarity is unable to sustain their weight.
The volume of the drop presented on
figures \ref{Figure06}-(c-e) is actually larger than $\mathcal{V}_{max}$:
these three images
represent the same drop at different times.
Their shapes cannot be understood solely with static arguments
and are affected by viscous stresses.

The exact value of $\mathcal{V}_{max}$ or,
in more general terms, the maximum value of the energy $E$
for a pendant drop depends in a subtle way
on both the geometry of the sustaining solid
and the imposed condition (such as fixed drop volume
or fixed pressure at a given altitude, for instance).
For instance, as shown in a detailed analysis
by Majumdar and Michael~\cite{DHMICHAEL},
a two-dimensional drop anchored to two parallel edges
located at the same altitude
destabilizes through three-dimensional perturbations
at a smaller volume than it would through two-dimensional ones
if the pressure is held constant at the anchoring edges. 
Another contribution to the stability of capillary surfaces can be 
found in \cite{GILLETTE}.

Conversely, the stability threshold for both types of perturbations
is the same (and hence higher) if the volume of the drop is fixed.

\subsection{Three-dimensional drops.}
\label{3Ddrops}
To conclude this section dedicated to 2D drops
(whose shape is invariant in one direction),
let us mention that
usual, 3D drops exhibit axial symmetry
and the azimutal curvature of the surface cannot be neglected in general.
In the case of a drop on a solid,
equation~(\ref{drop-pendulum1}) is changed into:
\begin{equation}
    \frac{d^2\theta}{ds^2}=-\sin\theta+
    \frac{d}{ds}\left(\frac{\sin\theta}{x_{\Sigma}/a}\right),
    \label{exactdrop}
\end{equation}
where $x_{\Sigma}$ is the horizontal distance of the current point
to the axis of symmetry of the drop
and $a$ is still the capillary length.
This equation clearly reduces
to equation~(\ref{drop-pendulum1}) for large values
of $x_\Sigma/a$,
{\it i.e.}, for the edge of a puddle (which is essentially
the same as that of a 2D puddle).
For small drops ($x_\Sigma/a<1$), the 2D equation we have used
is only in qualitative agreement with the observed shape.
In the same way, the exact equation for an axisymmetrical suspended drops is:
\begin{equation}
    \frac{d^2\theta}{ds^2}=-\sin\theta-
    \frac{d}{ds}\left(\frac{\sin\theta}{x_{\Sigma}/a}\right).
    \label{exactdrop1}
\end{equation}
A detailed study of these axisymmetrical capillary surfaces is 
presented in \cite{ORR}.

 \section{Flexible rod}
  \begin{figure}
    \centering
    \includegraphics*[scale=0.7]{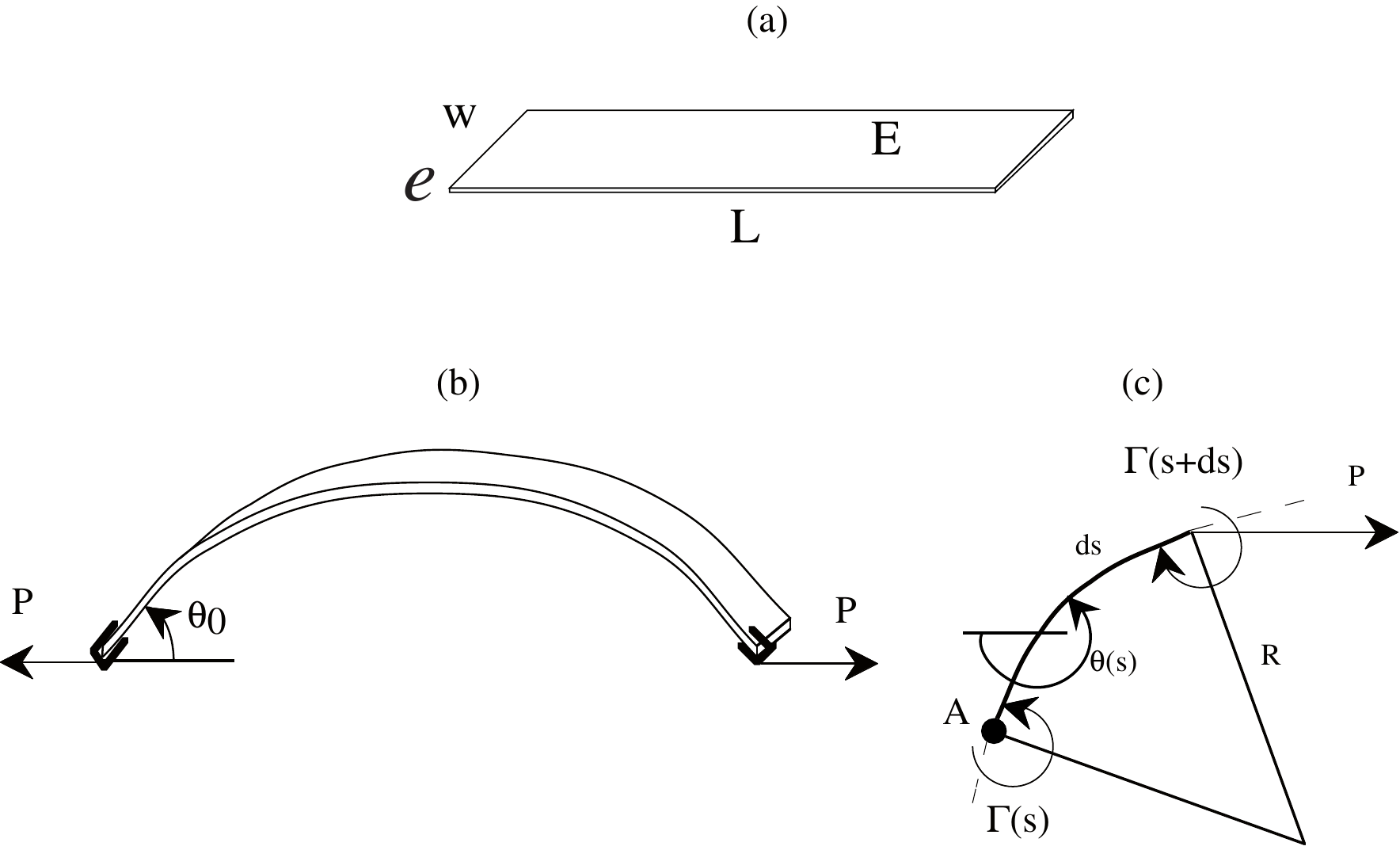}
    \caption{Experiment with a rod:
    (a) characteristic parameters of the rod,
    (b)  {rods subject to extension (outbound forces $P$)},
    (c) local torque balance  {and angle definition}.}
     \label{Figure08}
 \end{figure}

We now move to the flexible rod problem, depicted on
figure~\ref{Figure08}-(a). The rod is characterized by its length,
$L$, thickness, $e$, width, $w$, and Young modulus, $Y$.
Introducing the moment of inertia $I\equiv w e^3/12$ with respect
to the mid-plane, we now consider the rod as a massless,
inextensible line of bending rigidity $\mu\equiv YI$. Here again
we will study two cases (rod under tension or compression) with
two different geometric conventions.

\subsection{Rod subjected to a traction force $P$.}

\begin{figure}
    \centering
    \includegraphics*[scale=0.7]{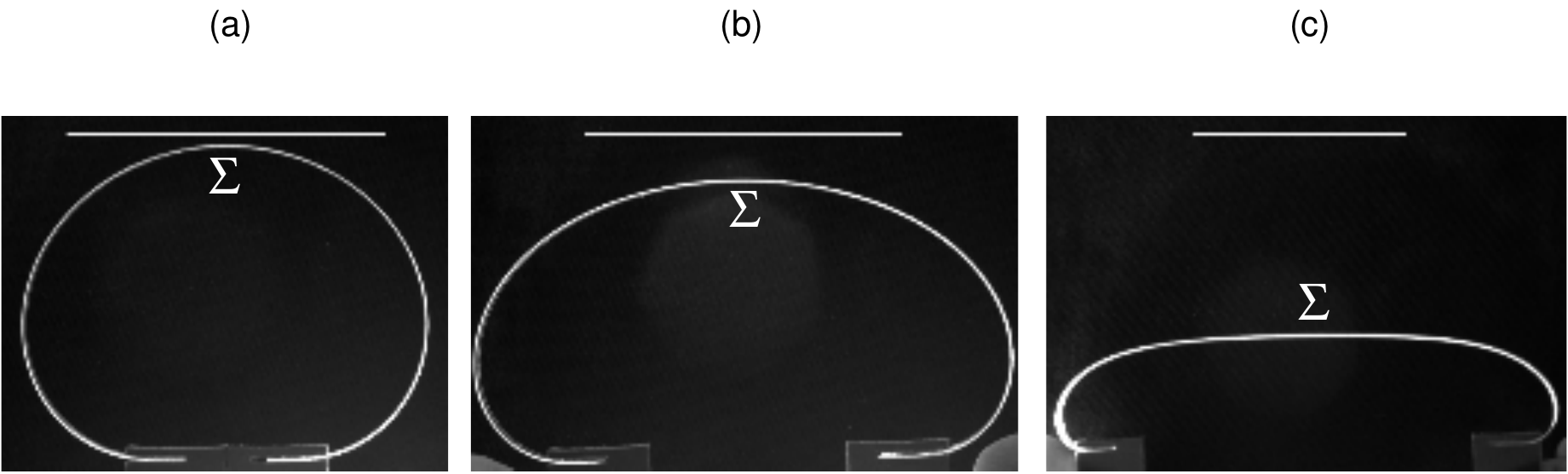}
    \caption{Examples of extended rods (increasing extension
      from a to c) obtained with
    $\theta_{0}=\pi$ (the horizontal bars are 8 cm long).}
\label{Figure09}
\end{figure}

Let us first consider that the rod is clamped
at both ends in symmetrical stirrups.
Let us impose some initial angle, $\theta_{0}$,
and a horizontal force, $P$,
as depicted on figure~\ref{Figure08}-(b).
Figure~\ref{Figure09} presents the  {shape} 
of a polycarbonate sheet
($\mu=5.10^{-4}\mbox{kg.m$^3$.s$^{-2}$}$)
 {subjected} to  {various} forces $P$
 {(increasing from left to right)}
with $\theta_{0}=\pi$.
These shapes are similar to those obtained with Mercury drops on a
solid (see figure~\ref{Figure03}). According to d'Arcy approach  {\cite{OGAF}}, this
suggests that they both share the same equation. Since the rod is
not subjected to any body forces (negligible weight), the local
force balance on a small element $ds$ of the rod is trivial:
$P(s)=P(s+ds)=P$. The torque balance, however, is not. With the
conventions  {of figure~\ref{Figure08}-(c)}
(note that $\theta$ is defined exactly as in the case
of a resting drop in figure~\ref{Figure02}-(a)), 
it can be written as:
 {
\begin{equation}
    \Gamma(s+ds) - \Gamma(s) +Pds\sin\theta=0,
    \label{rod01}
\end{equation}
where $\Gamma(s)$ is the  torque applied by the  {right-hand side} part of the rod 
(abscissa larger than $s$) onto the  {left-hand side} part of the rod (abscissa smaller than $s$). 
Its  value was related, by Euler \cite{Landau7},
to the radius of curvature $R$, through the relation $\Gamma(s)=\mu/R(s)=\mu d\theta/ds$.
}
Equation~(\ref{rod01}) thus reads:
\begin{equation}
    \frac{d^2\theta}{ds^2}=-\sin\theta,
    \label{rod02}
\end{equation}
where the characteristic length
\begin{equation}\label{arod}
a_{e}\equiv \sqrt{\mu/P}
\end{equation}
has been used for the non-dimensionalization.
The energy of the rod trajectory is given by:
\begin{equation}\label{enerr}
E=\frac12\dot{\theta}^2-\cos\theta
\end{equation}
(see equations~(\ref{enerp}) and~(\ref{enerm}) for comparison).
By symmetry $\theta_{\Sigma}=\pi$ at the middle point, $\Sigma$,
and the energy can be evaluated as $E=1+\dot{\theta}_{\Sigma}^2/2$.
 {Rods subject to outbound forces $P$ (as in figure~\ref{Figure08}-(b))} are thus
confined in the high energy region, $E_{>1}$.
Moreover, according to
equation~(\ref{rod02}), $|\dot{\theta}|$ reaches an extremum when
$\theta=\pi$. The puddle-like shape
shown on figure~\ref{Figure09}-(c)
indicates that this extremum is a minimum. The trajectories
of extended rods are thus symmetrical with respect to $\theta=\pi$
modulo $2\pi$.
They are strictly analogous to the drops on a solid,
the capillary length being replaced by the elastic length
$a_{e}$, and the contact angle by the clamping angle.
It can be observed that this length can be changed at will,
by varying either $\mu$ or $P$.
This is not the case with liquids on  {Earth,
since} $a$ is typically confined within the range $1$~mm~--~$5$~mm.

\subsection{Rod submitted to a compressive force P.}

\begin{figure}
    \centering
    \includegraphics*[scale=0.7]{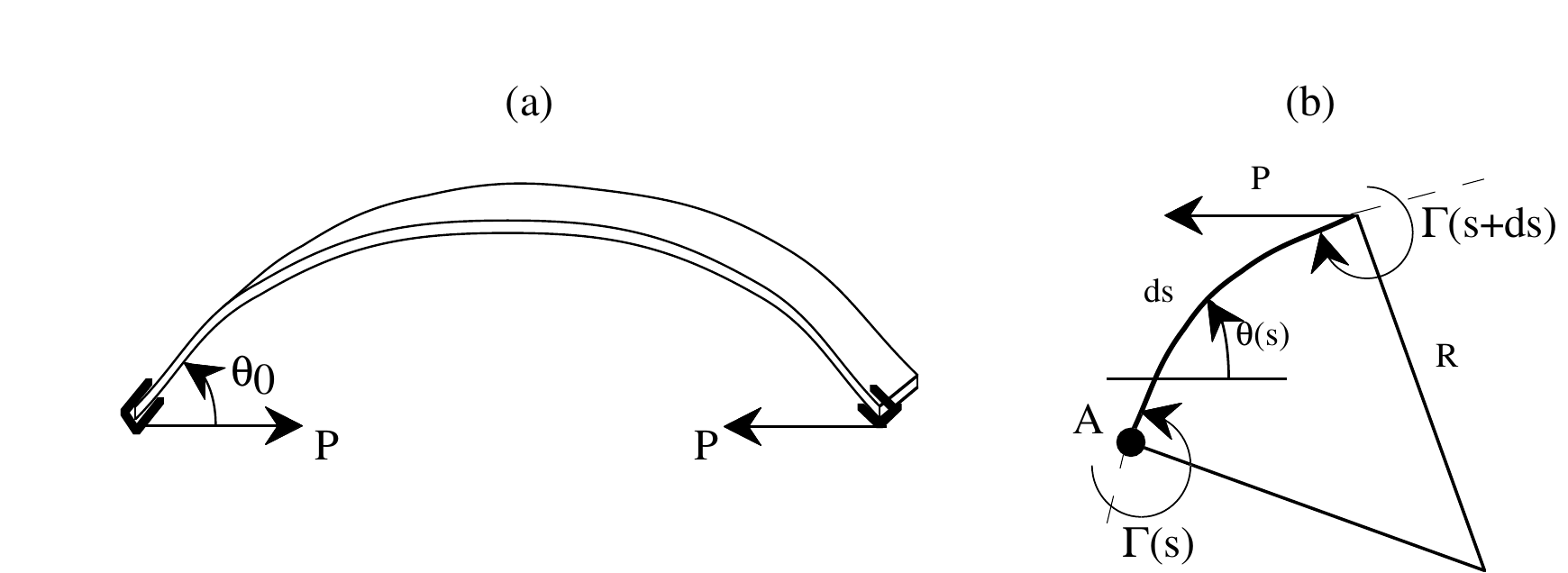}
    \caption{Compressed rod:
    (a) sketch of the experiment,
    (b) local torque balance.}
\label{Figure10}
\end{figure}
\begin{figure}
    \centering
    \includegraphics*[scale=0.7]{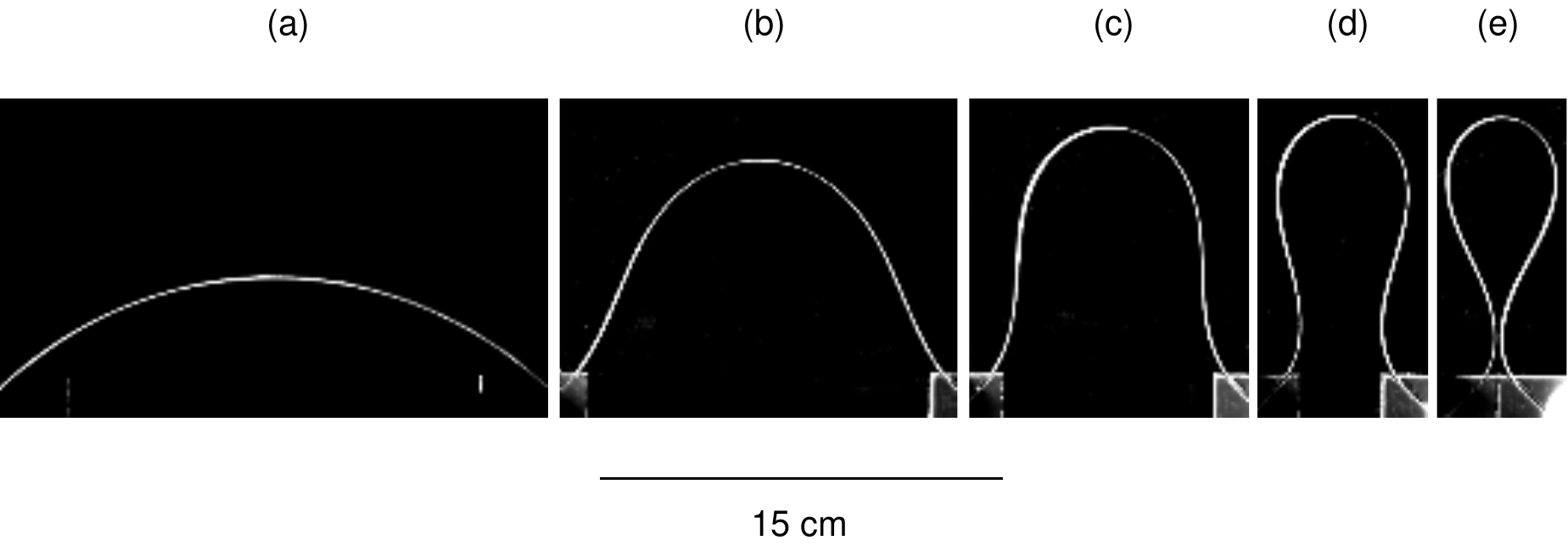}
    \caption{Examples of compressed rods obtained with
    $\theta_{0}=\pi/4$, the compression increases from (a) to (e).}
\label{Figure11}
\end{figure}

The clamped rod is now compressed by a force $P$
(see figure~\ref{Figure10}).
The shapes observed experimentally with
the polycarbonate sheet ($\mu=5.10^{-4}\mbox{kg.m$^3$.s$^{-2}$}$)
are presented on figure~\ref{Figure11} when subjected
to an increasing compression, keeping all other parameters constant.
These shapes are similar to those obtained with suspended drops
(figure~\ref{Figure06}). With the conventions  of
figure~\ref{Figure10}-(b),  {which are the same as those for
the pendant drop in figure~\ref{Figure02}-b,} the torque balance reads:
 {
\begin{equation}
    \Gamma(s+ds)-\Gamma(s)+Pds\sin\theta=0,
    \label{rod03}
\end{equation}
}
which yields the pendulum equation~(\ref{rod02}) upon
using the elastic length $a_{e}$ and the Euler relation
 {
$\Gamma=\mu\, d\theta/ds$.}
At the middle point, $\dot{\theta}$ is now a maximum,
which means that the trajectory is symmetric with respect to
$\theta=0$ modulo $2\pi$.
The inflection points, observed for example on figure~\ref{Figure11}-(e),
show that we have now access to the region $E_{<1}$:
compressed rods are thus strictly similar to suspended drops.
The forbidden region resulting from the gravitational constraint does
not exist  {in this case}.

  \section{ {Menisci}, looping and soliton}\label{MLS}
 \begin{figure}
    \centering
    \includegraphics*[scale=0.7]{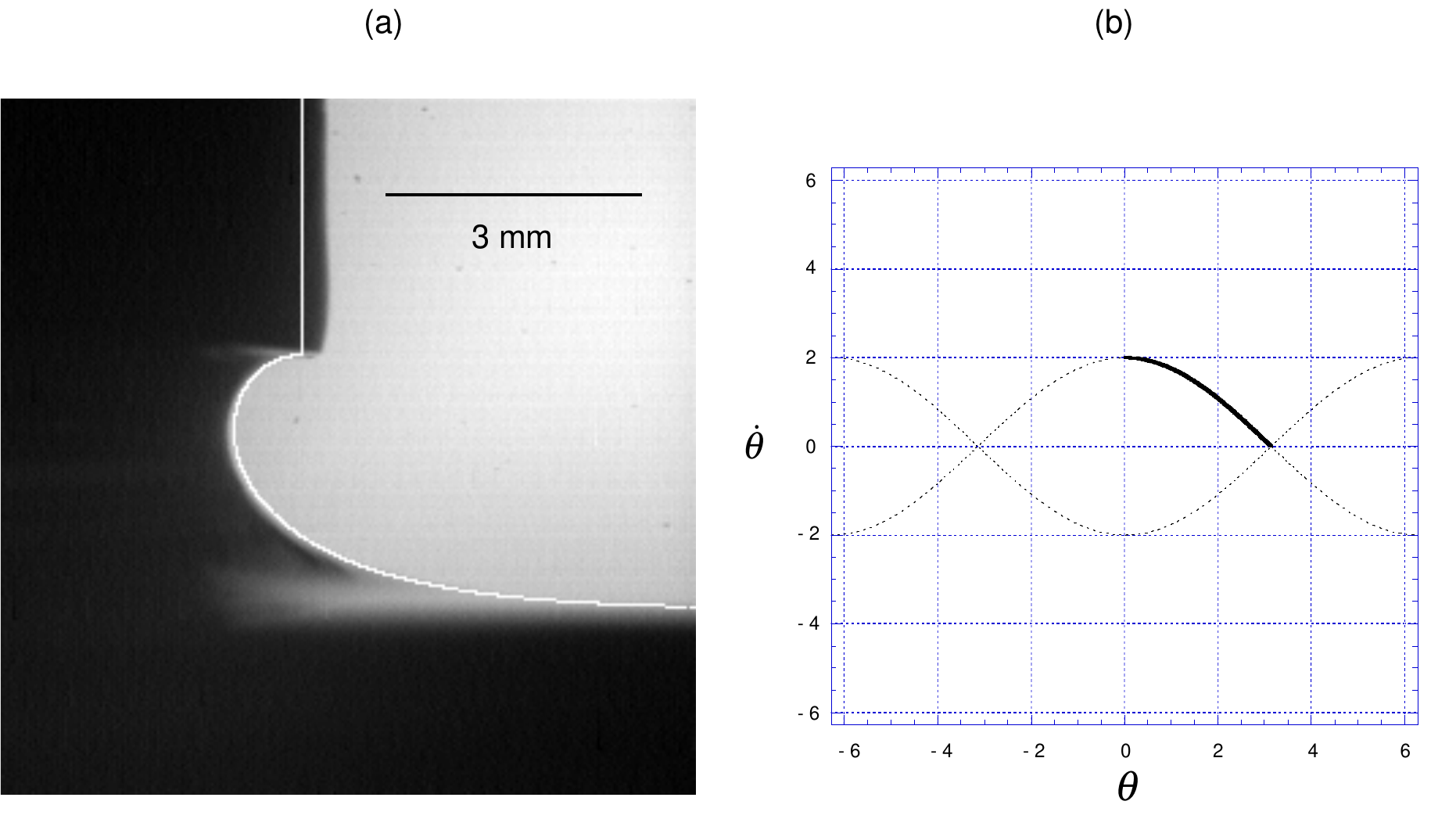}
    \caption{Shape of the meniscus:
    (a) comparison between the algebraic equation~(\ref{loop}) and
    a meniscus of silicone oil V1000, located under a
    column of large cross-section (diameter $D=20\mbox{ mm}$),
    (b) corresponding trajectory in the phase diagram
    $[\theta,\dot{\theta}]$.}
     \label{Figure12}
 \end{figure}

In the previous sections we have focused our attention
on both generic behaviors of all three systems,
which are characterized by closed orbits (energy $E<1$)
or open trajectories ($E>1$) in the $[\theta,\dot{\theta}]$ diagram.
To illustrate the analogy more deeply,
we now describe the particular behavior
that corresponds to the separatrix, $E=1$.

\begin{figure}
    \centering
    \includegraphics*[scale=0.7]{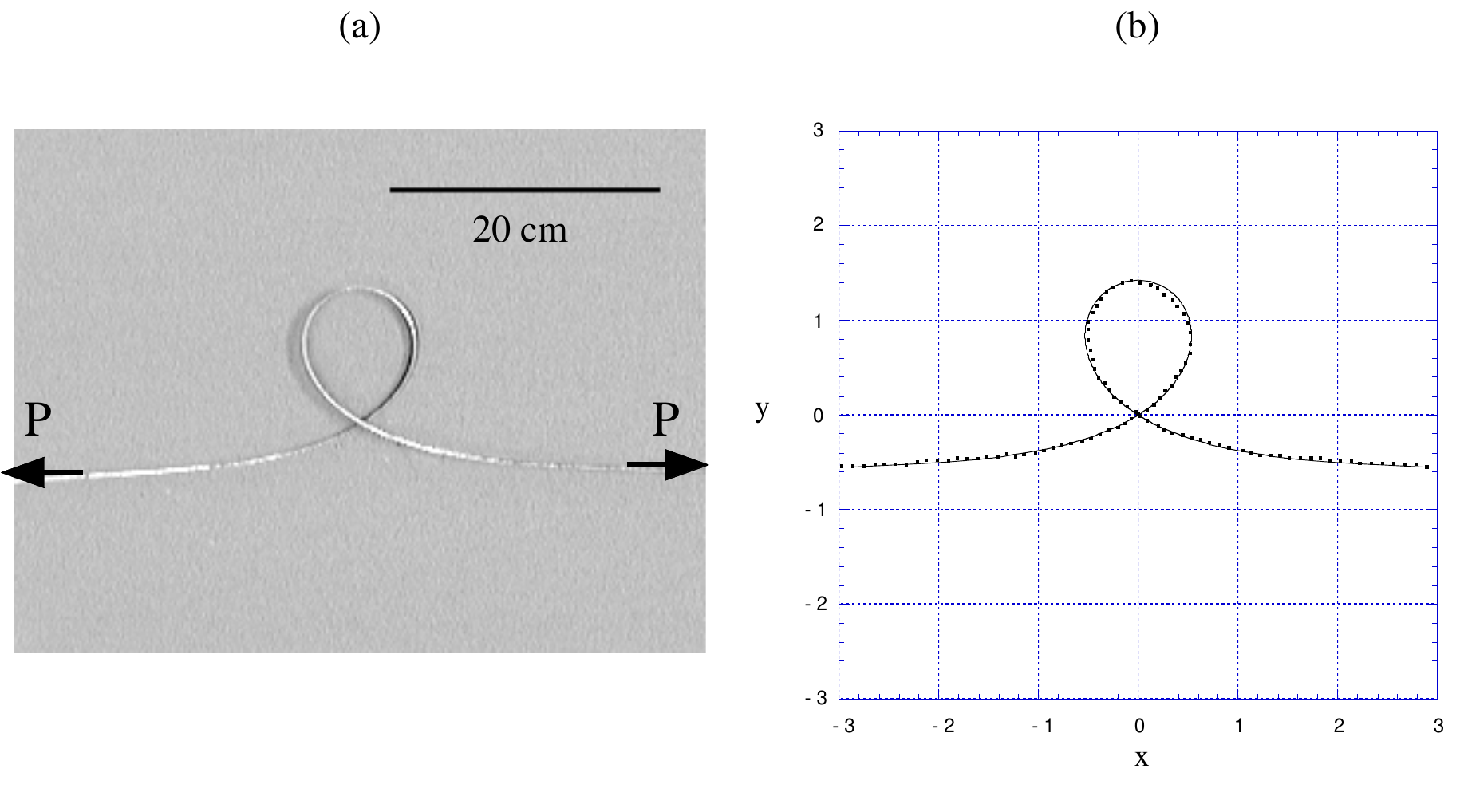}
    \caption{Meniscus-shaped elastica:
    (a) Stainless steel saw blade.
    (b) Experimental shape (dots)
      and plot (full curve) of the algebraic equation~(\ref{loop}).}
\label{Figure13}
\end{figure}

As opposed to the rotation motion ($E>1$)
or the oscillating regime ($E<1$),
the particular motion of the pendulum obtained
when it is dropped with vanishing velocity
from its unstable equilibrium position
is not periodic in time:
the pendulum slowly accelerates, then passes through its lowest position
at full speed and moves up again slowly,
until it finally stops at its uppermost position again.
On the $[\theta,\dot{\theta}]$ diagram,
this trajectory has two ends located in a bounded region,
which is not the case for all other trajectories.
Moreover, most of the  {motion} 
occurs during a finite time interval.
This particular motion is therefore a solitary excitation of the pendulum:
it is sometimes called the {\it soliton} motion,
although this term usually rather designates
localized, travelling waves that exist in many non-linear media.

The corresponding shape for a liquid
is the most usual occurence of a meniscus,
namely the curved region of the free surface
in the vicinity of the container wall,
see figure~\ref{Figure02}-(c).
It results from the fact that because of the local
surface tension balance,
the free surface has to meet the solid surface
with a fixed angle (known as the contact angle, $\theta_{e}$).
The dimension of the container is usually
much larger than the capillary length $a$,
which implies that both $\theta-\pi$
and the curvature $\dot{\theta}$ vanish far from the walls.
This indeed yields $E=1$ according to equation~(\ref{enerm}).
This equation can be integrated exactly in this particular case:
\begin{equation}
\sin\frac{\theta}{2}=\tanh s
\end{equation}
where the origin of the curvilinear distance
is taken at the apex of the looping ($\theta=0$).
It can also be integrated exactly
in cartesian coordinates~\cite{Landau6}:
\begin{equation}
   \frac{1}{2}\ln\left(\frac{1+\sqrt{1-y^2/4}}{1-\sqrt{1-y^2/4}}\right)
   -2\sqrt{1-y^2/4}=x.
    \label{loop}
\end{equation}
This function is plotted on figure~\ref{Figure12}-(a) and compared to
a meniscus of silicone oil V1000 located under the edge of a
column of diameter $D=20\mbox{ mm}$. The meniscus is essentially
two-dimensional since the column diameter is much larger than the
capillary length. The corresponding trajectory in the phase
diagram is presented in figure~\ref{Figure12}-(b).

The corresponding elastic conformation is that of
a very long flexible rod with one single curl,
loaded by a pure force at each end (no applied torque).

On figure~\ref{Figure13}-(a), we present a meniscus shape
formed with a stainless steel saw blade
($\mu=5.10^{-2}\mbox{kg.m$^3$.s$^{-2}$}$),
submitted to a traction force $P=10 \mbox{ N}$
(which corresponds to $a_{e}\approx 7\mbox{ cm}$).
On figure \ref{Figure13}-(b),
this looping is shown to be in nice agreement
with the algebraic equation~(\ref{loop}).

  \section{How deep is the  {analogy?}}

\begin{figure}
    \centering
        \includegraphics*[scale=0.7]{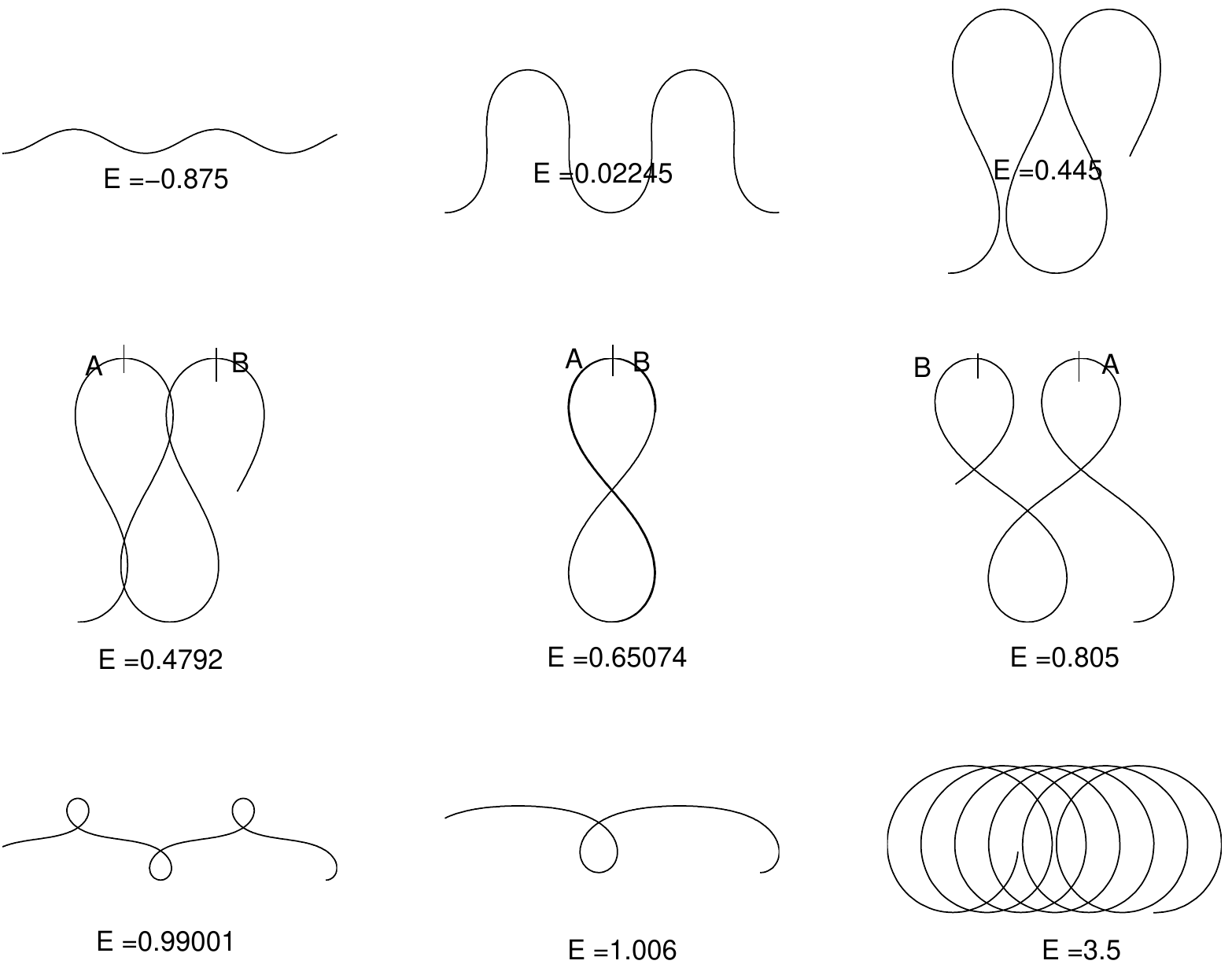}
    \caption{Various possible shapes obtained
from the pendulum equation
for different values of the pendulum energy $E$.
Some of these shapes can be materialized
with a meniscus or an elastica.
(a) $E$ slightly larger than $-1$ corresponds to gentle undulations;
(b) for $E\simeq 0.02245$, vertical slopes are reached;
(c-d) for $E\approx 0.46$, the curve now intersects itself:
an elastica needs to be thin enough (thread)
for crossings not to distort it too much in the third direction;
(e) for $E\approx 0.65074$, the curve is continuously self-superimposed (8-shape);
(f) for values of $E$ larger than about $0.65074$,
loops are formed; correlatively,
if the conventions for plotting the curve are preserved,
the overall motion along the curve is reversed as compared to smaller values of $E$,
as seen from the positions of points $A$ and $B$;
(g) as $E$ is further increased, the loops are located
further apart and are still alternatively pointing up and down;
(h) beyond the soliton solution $E=1$ for which only one looping is present,
the loops are now all pointing in the same direction;
(i) for high values of $E$,
successive loopings come close together and intersect.}
 \label{Figure14}
 \end{figure}

In the preceeding sections,
we presented the analogy between all three systems
essentially through the equation that describes them.
Figure~\ref{Figure14} illustrates the variety
of shapes that can be obtained in principle.

In this section, in order to
show a more precise physical correspondence between the
different actors in these analogies,
we first present a variational approach of each system,
which shows that although the equation of motion is the same,
the energy is somewhat different.
We then discuss an instability which is similar
for the meniscus and for the flexible rod.
We present the pendulum counterpart of the instability,
which displays some dissimilarities with the other two.
We finally briefly study the behavior of a flexible membrane
enclosing a liquid, which interpolates between
the elastic rod and the meniscus.
 {The table in figure~\ref{Table01} }summarizes several aspects
of the analogy.\newline

\begin{figure}
{\scriptsize \centering
\begin{tabular}{|c|c|c|c|}
 \hline
   & Pendulum
  & Drops & Flexible rod
  \\  (Eqs.) in text  &
  &  &
  \\ \hline
  parameters &
  \begin{tabular}{c}
     angle $\theta$\\
     time $t$\\
     velocity $l\dot{\theta}$\\
  \end{tabular} &
  \begin{tabular}{c}
     angle $\theta$\\
     curv. distance $s$\\
     curvature $\dot{\theta}$\\
  \end{tabular} &
  \begin{tabular}{c}
     angle $\theta$\\
     curv. distance $s$\\
     curvature $\dot{\theta}$\\
  \end{tabular}
  \\ \hline
 \begin{tabular}{c}
     characteristic \\scale\\
    (\ref{Eq1}) (\ref{aliq}) (\ref{arod}) \\
  \end{tabular}&
  \begin{tabular}{c}
     $\tau=\sqrt{\frac{l}{g}}$\\
     length $l$, gravity $g$\\
  \end{tabular} &
  \begin{tabular}{c}
     $a=\sqrt{\frac{\sigma}{\rho\,g}}$ \\
     surface tension $\sigma$ \\
     density $\rho$\\
  \end{tabular} &
  \begin{tabular}{c}
     $a_e=\sqrt{\frac{\mu}{P}}$ \\
     applied force $P$ \\
     bending rigidity $\mu$\\
  \end{tabular} \\
  \hline
  \begin{tabular}{c}
    simplest \\
    governing eq. \\
    (\ref{eqp1}) (\ref{equality}) (\ref{rod3a}) \\
  \end{tabular}  &
  \begin{tabular}{c}
     angular acceleration \\
     due to gravity \\
     $ml\ddot{\theta}=-mg\sin\theta$\\
  \end{tabular} &
  \begin{tabular}{c}
     $\sigma\dot{\theta}=-\rho gy$ \\
     pressure\\
  \end{tabular} &
  \begin{tabular}{c}
     $\mu\dot{\theta}+Py=0$ \\
     torque \\
  \end{tabular}
 \\ \hline
  \begin{tabular}{c}
    conserved \\
    "energy" $E$  \\
    (\ref{enerp}) (\ref{enerm}) (\ref{enerr}) \\
  \end{tabular} &
  $\frac12ml^2\dot{\theta}^2-mgl\cos\theta$ &
  $\frac12\rho gy^2-\sigma\cos\theta$ &
  $\frac12\mu\dot{\theta}^2-P\cos\theta$
\\ \hline
  \begin{tabular}{c}
  $E<1$\\ \\ $E>1$\\
  \end{tabular}
 &
  \begin{tabular}{c}
  oscillations\\ \\ rotations\\
  \end{tabular}
 &
  $\left\} {\rm
  \begin{tabular}{c}
     Drops and bubbles \\ bounded by \\
     angle with \\ solid surface\\ \\
  \end{tabular} }\right.$ &
  \begin{tabular}{c}
  compressed\\ \\ stretched\\
  \end{tabular}
\\
  $E=1$&  soliton &
  2D meniscus &
  single looping
 \\ \hline
 unperturbed state &  \begin{tabular}{c}
  $\theta(t)=\pi$ \\
   (unstable equil.) \\
  \end{tabular} &
  \begin{tabular}{c}
  $\theta(s)=0$ \\
  (liquid above air) \\
  \end{tabular} &
  \begin{tabular}{c}
  $\theta(s)=0$ \\
  (compressed) \\
  \end{tabular}
 \\
 perturbation &
 angular frequency $\omega$ &
  wavenumber $k$ &
  wavenumber $k=2\pi/L$
 \\
  instability &  {}  &
  Rayleigh-Taylor &
  Euler buckling
\\
  threshold &
$\omega.\tau<{\cal O}(1)$ &
  $k.a<{\cal O}(1)$ &
  $k.a_e<{\cal O}(1)$
\\ \hline
\end{tabular}\newline\newline
\caption{ {Comparison of the different systems} \label{Table01} } }
\end{figure}

 \subsection{Variational description}

Comparing the governing equation, as we have done so far, shows a
deep similarity between all three systems. Among them, the
meniscus and the elastic rod are most similar since the variable
$s$ has the same spatial meaning: we now turn to a variational
approach of these two systems, in the case of the pendant drop,
and of the compressed rod for simplicity.

For the liquid, the following quantity needs
to be minimized:
 \begin{equation}\label{energyliq}
 \int_0^X{\cal L}(y,y^\prime)\,{\rm d}x,
 \hs{\rm where}\hs
 {\cal L}=-\frac12\,\rho\,g\,(y-y_0)^2
 +\sigma\,\sqrt{1+y^{\prime\,2}},
 \end{equation}
where the first term is the gravitational potential energy of the
liquid column located above each point of the interface with
respect to some reference altitude $y_0$ and where the second term
is the interfacial energy.
 The full Euler-Lagrange equation reads:
 \begin{equation}\label{men1a}
\frac{\sigma}{R} + \rho\,g\,(y-y_0)=0,
 \end{equation}
where $1/R=\dot{\theta} =y^{\prime\prime}/(1+y^{\prime\,2})^{3/2}$
is the  surface curvature. This is relation was found in section
(3.2). Upon differentiation with respect to $s$
 we again find the pendulum equation~(\ref{rod02}):
 \begin{equation}\label{men2a}
\sigma\,\ddot{\theta}+\rho\,g\,\sin\theta=0
 \end{equation}
 A first integral can be obtained
 as $y^\prime\partial{\cal L}/\partial y^\prime
 -{\cal L}={\rm const}$:
 \begin{equation}\label{men3a}
 \frac12\,\rho\,g\,(y-y_0)^2
 - \sigma\cos\theta={\rm const}
 \end{equation}
(where $\cos\theta=1/\sqrt{1+y^{\prime\,2}}$). When combined with
equation~(\ref{men1a}), this yields the energy (see
equation~(\ref{enerm}) for comparison):
\begin{equation}\label{men4}
E=\frac12 {\dot{\theta}}^2-\cos\theta
\end{equation}
where $s$ has been rescaled by $a=\sqrt{\sigma/(\rho g)}$.\newline

The rod energy per unit length is $\frac12 \mu\dot{\theta}^2$.
Enforcing a prescribed value for the horizontal distance
$\int\cos\theta\,{\rm d}s$
swept by the entire rod of length $L$,
yields an additional term in the energy to be minimized:
 \begin{equation}\label{energyrod}
 \int_0^L{\cal L}(\theta,\dot{\theta})\,{\rm d}s
 \hs{\rm where}\hs
 {\cal L}=\frac12\,\mu\dot{\theta}^2
 +P\cos\theta
 \end{equation}
where the Lagrange multiplier $P$ is the force
(assumed to be horizontal) applied at both ends of the rod.
The above integral can be interpreted as the action,
where the first term in ${\cal L}$ is the kinetic energy
and the second term is the opposite of the potential energy.
The corresponding full Euler-Lagrange equation
$(d/ds)[\partial{\cal L}/\partial\dot{\theta}]
-\partial{\cal L}/\partial\theta=0$
is the pendulum equation (see equations~\ref{rod02}
and~\ref{men2a}):
 \begin{equation}\label{rod1a}
 \mu\,\ddot{\theta}+P\sin\theta=0
 \end{equation}
 But since ${\cal L}$ does not depend
 explicitely on $s$,
 a first integral can be obtained directly
 as $\dot{\theta}\partial{\cal L}/\partial\dot{\theta}
 -{\cal L}={\rm const}$:
 \begin{equation}\label{rod2a}
 \frac12\,\mu\,\dot{\theta}^2
 -P\cos\theta={\rm const}
 \end{equation}
which is the energy of the rod trajectory (equation~\ref{enerr}),
just like equation~(\ref{men4}) is that of the meniscus trajectory.
Alternatively, integrating the pendulum equation
with respect to the curvilinear distance $s$
leads to the condition of local torque balance:
\begin{equation}\label{rod3a}
\mu\,\dot{\theta}+P\,(y-y_0)=0
\end{equation}
(where the torque is assumed to vanish at altitude $y_0$),
which is the same as equation~(\ref{men1a}).\newline

The present variational approach
shows that the energy of the meniscus
 is a natural function of the altitude $y$,
while that of the rod is expressed
in terms of the angle $\theta$.
The reason for this discrepancy is essentially
that the energy of the rod is unchanged by translation,
whereas if the meniscus is translated vertically,
the gravitational energy of the liquid is altered.
 As a consequence, the corresponding
 Euler-Lagrange equations differ also
 (see equations~\ref{men1a} and~\ref{rod1a}).
 In fact, the difference between both problems
lies essentially in a derivative:
equation~(\ref{men2a}), which is the same
as the rod Euler-Lagrange equation~(\ref{rod1a}),
is also the derivative of the meniscus
Euler-Lagrange equation~(\ref{men1a}).
That is due to the fact that the angle $\theta$
is essentially the derivative of the altitude $y$
(more precisely, $\tan\theta={\rm d}y/{\rm d}x$
and $\sin\theta=-{\rm d}y/{\rm d}s$).

As a conclusion, although the equilibrium shape of these systems
obey the same equation, it is difficult to say that the underlying
physics are identical. The energies involved in the two problems
are not similar. There is a chance that their dynamics are
different. However we will see in next section that they display
similar instabilities.

\subsection{Rayleigh-Taylor Instability and Buckling Instability.}

The well-known Rayleigh-Taylor instability of a liquid and the
buckling instability of a compressed rod (due to Euler)
almost have a pendulum counterpart.
In the present paragraph, we review the first two.
We discuss in detail the analogous pendulum instability
in the next paragraph.

\begin{figure}
    \centering
    \includegraphics*[scale=0.7]{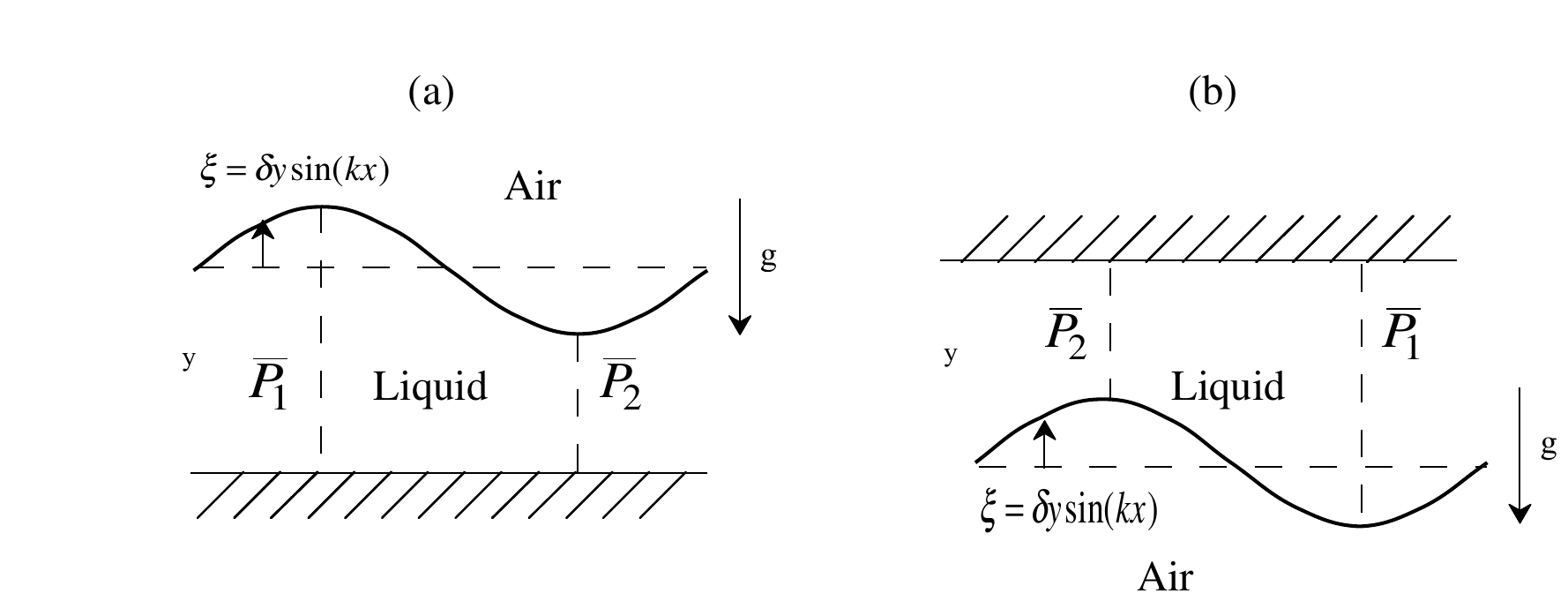}
    \caption{Rayleigh-Taylor Instability:
    (a) Stable situation when air is above the liquid.
    (b) Possible unstable situation when air is below the liquid.}
\label{Figure15}
\end{figure}

If the free surface of a liquid (assumed to be horizontal) is
submitted to a sinusoidal perturbation $\xi=\delta y\,\sin(kx)$, its
evolution will depend crucially on whether air is above
(\ref{Figure15}-(a)) or below the liquid (\ref{Figure15}-(b)).
Assuming the liquid is initially at rest,
a point located within the liquid
just beneath the free surface
will be accelerated due to the surface deformation:
$\ddot{\xi}\propto\mp\rho g\,\xi-\sigma k^2\,\xi$.
where the minus (resp., plus) sign corresponds to the situation of
figure~\ref{Figure15}-(a) (resp., b).
The first term corresponds to the hydrostatic pressure
exerted by the bulk of the liquid,
which is stabilizing in case (a)
and destabilizing in case (b).
The second term is the Laplace pressure
exerted by the curved free surface,
which is always stabilizing.
The evolution can be written as:
\begin{equation}
\ddot{\xi}\propto
\rho g\,\xi\,[\mp 1-(ka)^2].
    \label{rod04}
\end{equation}
Hence, when air is above the liquid,
the square bracket is always negative: puddles are stable.
In the reverse situation (figure~\ref{Figure15}-b),
the system is stable for wavelengths
smaller than $2\pi a$ but becomes unstable when:
\begin{equation}
k.a<{\cal O}(1) \label{instabm}
\end{equation}
Hence, pendant puddles of extension larger than $2\pi a$ are
unstable. This is the Rayleigh-Taylor instability~\cite{CH}.

Let us now turn to a straight rod which is either stretched or
compressed with a force $P$ that is aligned with it. The stretched
situation is  {analogous} to a drop resting on a solid and is thus
always stable. The compressed rod, however, is unstable (buckling
was probably first described by L.~Euler) when the force exceeds
the critical value $P_{c}=\mu k^2/b$, {\it i.e.}, when:
\begin{equation}
k.a_e<\sqrt{b} \label{instabr}
\end{equation}
where $k$ is the wavenumber of the perturbation (on the order of
the inverse rod length) and $b$ is a constant that depends on the
boundary conditions~\cite{Landau7}. The rod thus buckles when its
size is larger than the elastic length $a_{e}$. This is therefore
directly analogous to the Rayleigh-Taylor instability.\newline

 This might be surprising because the analogy we described
was restricted to equilibrium shapes
and instabilities are a true expression of the dynamics,
which are certainly very different.
 Indeed for a rod, the length of the line is always conserved, and the
 final state will be a buckled rod, but for an unstable liquid subject to the
 Rayleigh-Taylor instability, the length of the interface
 varies at fixed enclosed volume, and the final state is
 obtained when the liquid has fallen down.

However, the knowledge of all the equilibrium states as a function of a
 governing parameter is sufficient to understand the dynamics qualitatively.
 Before the instability is developed, the conserved quantities such as
the size $x$ of the liquid interface
(which ensures volume conservation)
 and the length $l$ of the rod
 are \emph{linearly} the same. Hence, neutral modes
(which are exactly like other equilibrium
 solutions) are identical in both systems,
and so are thus the instability criteria.

\subsection{Corresponding pendulum  {instability?}}
%

The pendulum does not display an instability
strictly similar to
the Rayleigh-Taylor instability for the meniscus
or to the Euler buckling instability for the rod
described in the last paragraph.

The first, obvious difference is that
the situation where the meniscus can display
the Rayleigh-Taylor instability (liquid above air)
and the situation where the rod can buckle (compressed state)
correspond, {\it via} the analogy,
to the bottom position of the pendulum,
which is unconditionally stable.

A further difference between the pendulum
and the meniscus or the rod
is that the curvilinear distance $s$ is now
turned into the time $t$:
the strict equivalent of a sinusoidal
perturbation (wavenumber $k$) of the rod or the liquid surface is
a sinusoidal movement of the pendulum at some angular frequency, $\omega$.
This is not transposable as such
because the pendulum has no other degree of freedom:
if its position is prescribed at
all times, the problem becomes pointless. 
This is in contrast with
the liquid surface (or the rod) whose shape can be prescribed at
$t=0$ and which can still evolve freely at later times. For the
pendulum, the time $t$ is already the equivalent of the spatial
coordinate $s$.
 {To date, we have not found any setup that would display an equivalent instabliity. Any suggestion is welcome.}

We note that the classical parametric forcing~\cite{L1}
yields another type of instability.
The attachment point of the pendulum
is vibrated vertically,
which amounts to oscillating the parameter $g$.
In this situation, a harmonic oscillator can be excited
by a signal at twice its own frequency
or at subharmonics thereof.
 {Such a behavior is very different
from the ones considered here (Rayleigh-Taylor and buckling)}.
Furthermore, the pendulum is non-linear
and its resonance frequency depends on the amplitude,
hence the frequency of a weak parametric forcing
should be adjusted dynamically in order to bring the pendulum to large amplitudes.

 \subsection{Flexible sheet enclosing a liquid}
%
 Earlier in this section, we showed
 that the energy of the rod and that of the meniscus
 cannot be written in the same fashion.
 Here, we consider a system
 that somehow interpolates between
 the meniscus and the rod.

 In the two-dimensional geometry, the free surface
 of the liquid transmits a force
 which is tangential and of constant amplitude,
 and upon bending, it is able to support
 the pressure of the liquid which is normal.
 By contrast, due to its bending rigidity,
 the rod transmits forces of arbitrary orientations
 and transmits also torques,
 but it is not subjected to any external stresses
 other than the end-loading forces and torques.

 In order to combine these features,
 we here consider a flexible membrane
 (with a finite bending rigidity $\mu$)
 with air on one side and a liquid on the other side.
 We suppose that the membrane is inextensible~\cite{inextensible}.

 The stresses in the membrane are then given by:
 \bee
 \dot{\Gamma}&=&-P.n\label{flm1}\\
 \dot{P}&=&p\,n\label{flm2}\\
 \Gamma&=&-\mu\dot{\theta}\label{flm3}\\
 \dot{p}&=&-\rho g\sin\theta\label{flm4}
 \eee
 where $P$ and $\Gamma$ are the usual force and torque
 transmitted along the membrane (curvilinear abscissa $s$),
 and where $n$ and $u$ are the unit vectors
 normal and tangent to the membrane, respectively.
 The first two equations express the local torque
 (see equation~\ref{rod01}) and force balances,
 the third equation reflects the bending elasticity of the membrane
 and the fourth gives the hydrostatic pressure in the liquid.

 {From} equations~(\ref{flm1}) and~(\ref{flm3}),
   the normal component of the transmitted force
   can clearly be expressed as
   \begin{equation}
       P.n=\mu\ddot{\theta}
   \end{equation}
   Using the derivatives of the unit vectors
   $\dot{u}=\dot{\theta}n$ and
   $\dot{n}=-\dot{\theta}u$,
   the tangent component of the force can be derived from
   $d(P.u)/ds=\dot{P}.u+\dot{\theta}P.n =0+\mu\ddot{\theta}\dot{\theta}$:
   \bee
        P.u=\frac12\mu\dot{\theta}^2-A
   \eee
 where $A$ is a constant
 (in the limit of vanishing bending rigidity,
 it is equivalent to the surface tension $\sigma$).
 From the expression of $P.n$:
 \bee
 \mu\theta^{(3)}&=&d(P.n)/ds
 =\dot{P}.n-\dot{\theta}P.u\\
 &=&p-\frac12\mu\dot{\theta}^3+A\dot{\theta}
 \label{memb3rd}
 \eee
 Finally, the shape of the membrane is described
 by a fourth order ordinary differential equation:
 \begin{equation}
 \mu\theta^{(4)}
 +\frac32\mu\ddot{\theta}\dot{\theta}^2
 -A\ddot{\theta}+\rho g\sin\theta=0
 \label{memb4th}
 \end{equation}
 The unknown parameter $A$
 and the solution to the equation
 are determined from five initial conditions at $s=0$:
 the angle $\theta$, the pressure $p$,
 the applied torque $\Gamma(0)=-\mu\dot{\theta}(0)$
 and both components of the applied force $P(0)$.
 \begin{eqnarray}
 A&=&\frac12\Gamma^2(0)/\mu-P(0).u(0)\label{thA}\\
 \theta(0)&=&{\rm specified}\label{th}\\
 \dot{\theta}(0)&=&-\Gamma(0)/\mu\label{thp}\\
 \ddot{\theta}(0)&=&P(0).n(0)/\mu\label{thpp}\\
 \theta^{(3)}(0)
 &=&p(0)/\mu+\Gamma(0)P(0).u(0)/\mu^2\label{thppp}
 \end{eqnarray}

 The full equation~(\ref{memb4th}) clearly reduces
 to the meniscus equation~(\ref{drop-pendulum1})
 in the limit of vanishing bending rigidity $\mu$:
 \begin{equation}
 -A\ddot{\theta}+\rho g\sin\theta=0
 \label{reducmen}
 \end{equation}

 In a region where the membrane spans only a small
 vertical distance, the pressure $p$ is essentially constant
 and the shape is described by equation~(\ref{memb3rd})
 which does not depend explicitely on $\theta$
 (because the problem is then invariant by rotation).
 If, furthermore, the pressure is negligible,
 then this equation reduces to:
  \begin{equation}
     \theta^{(3)}
 +\frac12\dot{\theta}^3-\frac{A}{\mu}\dot{\theta}=0
 \label{noliquid}
 \end{equation}
 One can check that the usual rod conformation
 (see equation~\ref{enerr}) specified
 by the initial condition~(\ref{th}) and by:
 \bee
 \frac12\dot{\theta}^2
 +k\cos[\theta-\alpha]-\frac{A}{\mu}=0
 \label{reducrod}
 \eee
 is then duly the solution of equation~(\ref{noliquid})
 with the specified initial conditions,
 provided that $k$ and $\alpha$ are chosen
 in such a way that:
 \bee
 -k\cos[\alpha-\theta(0)]&\equiv&P(0).u(0)/\mu\\
 -k\sin[\alpha-\theta(0)]&\equiv&P(0).n(0)/\mu
 \eee
 In other words, $\alpha$ is the oriented angle
 between the tangent vector $u$
 and the applied force $P$, and $k=-|P|/\mu$.

\begin{figure}
    \centering
        \includegraphics*[scale=0.7]{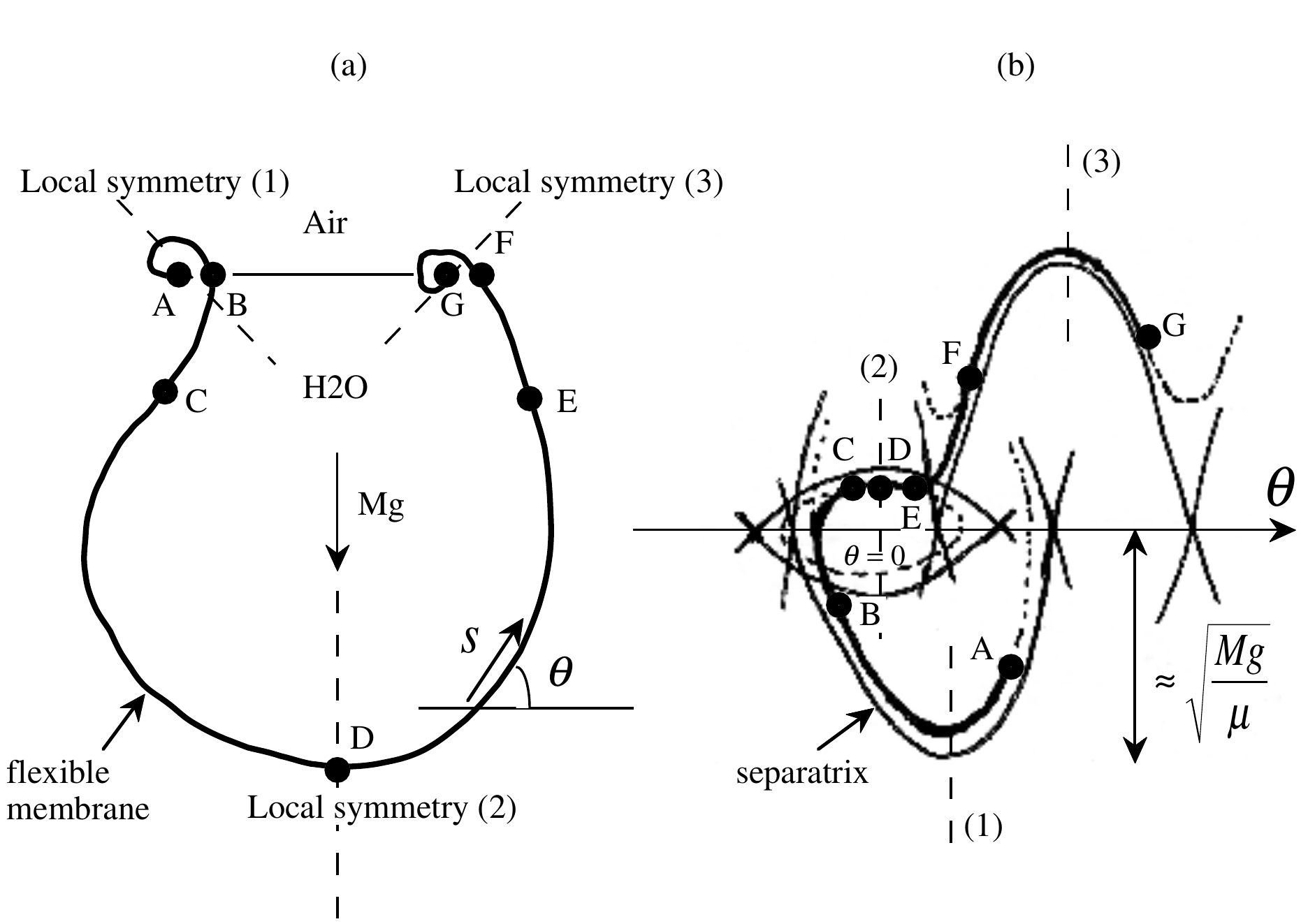}
    \caption{(a) Inextensible, flexible membrane
         with finite bending rigidity enclosing a liquid.
         (b) Trajectory of the flexible membrane
         in the $[\theta,\dot{\theta}]$ diagram.
         It interpolates between a meniscus-like region
         (with a large characteristic length-scale)
         and a rod-like region near the clamped ends
         of the membrane (with a small length-scale).}
 \label{Figure16}
 \end{figure}

 In general, the full equation~(\ref{memb4th})
 does not have pendulum-like solutions:
 such a solution is only valid locally.
 For instance, suppose that the flexible membrane
 is clamped on both ends at the same altitude,
 at a fairly small distance from one another,
 and that the liquid is poured into the resulting
 hollow-shaped membrane (see figure~\ref{Figure16}-(a)).
 On the large scale, the curvature of the membrane is small
 and if the bending rigidity $\mu$ is not too large,
 the shape is simply that of a pendant drop,
 according to equation~(\ref{reducmen}).
 Near the clamped ends of the membrane, however,
 depending on the orientation that is imposed,
 the curvature can be quite high. 
 Since the pressure of the liquid is negligible there,
 we recover a rod-like conformation,
 see equation~(\ref{reducrod}).
 The characteristic length scales of these two regions
 of the membrane are not the same,
 so that on a $[\theta,\dot{\theta}]$ diagram,
 the trajectory of the membrane goes from
 a first pendulum-like solution with one length scale
 to a second such solution with another length scale
 (see figure~\ref{Figure16}-(b)).
 In the crossover region, both the membrane bending rigidity
 and the liquid pressure are important
 and the full equation~(\ref{memb4th}) does not reduce
 to a pendulum equation.

\begin{figure}
    \centering
  \includegraphics*[scale=0.7]{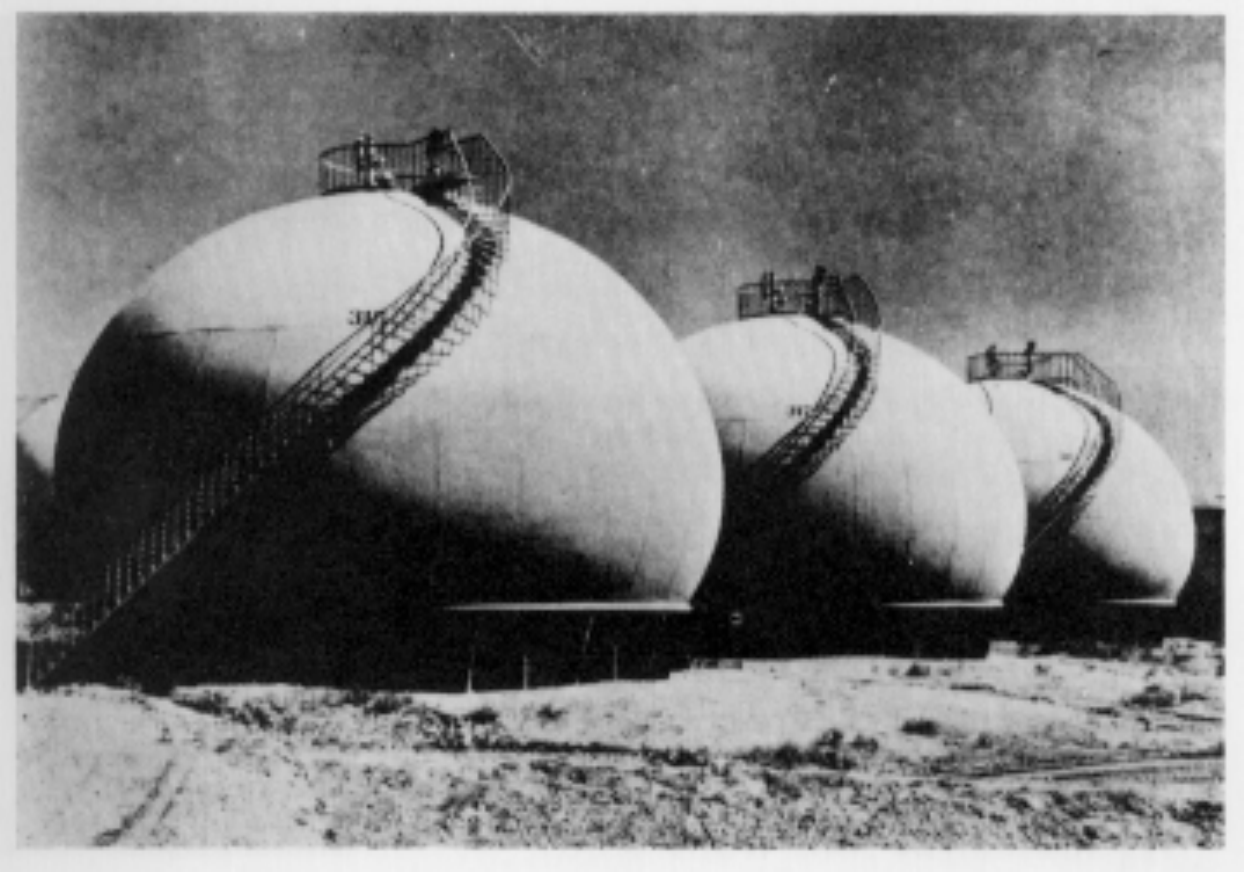}
    \caption{Storage tanks with metallic staircases for maintenance
         (from McMahon and Bonner,
         {\it On Size and Life}). They are called Hortonspheroids
         and generally contain hydrocarbons under pressure.
         Their shape closely follows that of liquid drops
         such as those presented on figure~\ref{Figure03},
         although discrepancies are due to other real engineering
         issues such as the transition of forces at the interface
         with the foundation.
         {\it Photo and comments:
         courtesy of Chicago Bridge \& Iron Company.}}
 \label{Figure17}
 \end{figure}

For special initial conditions, however,
 even the full equation~(\ref{memb4th})
 has a pendulum-like solution.
 Indeed, one can show that the trajectory defined
 by the initial condition~(\ref{th}) and by:
 \bee
 \frac12\dot{\theta}^2
 +k\cos\theta+\frac{\rho g}{\mu k}-\frac{A}{\mu}=0
 \label{pendumemb}
 \eee
 is the solution of equation~(\ref{noliquid}) if:
 \bee
 &&k\cos[\theta(0)]\equiv -P(0).u(0)/\mu-\frac{p(0)}{\Gamma(0)}\\
 &{\rm and}&k\sin[\theta(0)]\equiv +P(0).n(0)/\mu
 \eee
 These two equations yield a unique value of $k$,
 provided that the initial conditions at $s=0$ obey the following:
 \bee
 P(0).n(0)\cos[\theta(0)]
 +\left\{P(0).u(0)+p(0)\mu/\Gamma(0)\right\}\sin\theta(0)=0
 \eee
 To conclude this section, let us mention that
 containers used to store large amounts of liquid 
  {under pressure}
 (see figure~\ref{Figure17})
 have essentially the same shape as liquid drops
 resting on a solid (figure~\ref{Figure03}),
 as pointed out by McMahon and Bonner~\cite{OSAL}
 in their beautiful book {\it On Size and Life}.
 This may seem surprising since the interface between
 the stored liquid and the air
 ({\it i.e.}, the container walls)
 is not precisely a free surface~!
 The steel walls can transmit forces and torques,
 and as was described above, the shape of the container
 could therefore be very different from the pendulum equation.
 In fact, the shape of the container
 ({\it i.e.}, in its initial, unloaded state)
 was designed to match that of a drop for a precise reason:
 at least when the container is full of liquid  {at the appropriate pressure},
 the forces are oriented tangentially within the walls~\cite{TANGENT}
 and their intensity is the same everywhere,
 just like surface tension.
 This very clever design thus requires walls
 of essentially constant thickness.
 Moreover, the absence of transmitted torques
 in a full container reduces the risks of
 additional wall bending and fatigue.
The  {bottom part of the} real shape, however, is slightly different
 {from that of a liquid drop}
due to various engineering issues
and in particular the need for an appropriate transmission of stresses
between the walls of the tank and the foundation.

 \section{Historical note}
The present work started independently in Marseille~\cite{TB} and in
Paris~\cite{LARECH},\cite{LARECH1},\cite{LARECH2} until two of the 
authors  met for dinner, one ocean away from their home country.
 They happened to talk about a knife, a glass of water
 and a pair of laces,
 and all at once they knew that they had been
 thinking about the very same analogy
 for several months already.
 They wish to dedicate this work
 to all those who certainly have also pondered
 this analogy over the years. 

 A similar analogy was drawn by G.~I.~Taylor
 and A.~A.~Griffith
 in a series of three papers~\cite{TORSION}.
 The height profile $h(x,y)$
 of a soap film attached to a planar contour
 and subjected to a small pressure difference
 obeys $(\partial_x^2+\partial_y^2)\,h={\rm const}$.
 The same equation is obeyed by a function
 which describes the shear stresses
 in the section of a twisted bar,
 the contour of the section being
 the same as that of the soap film.
 This analogy was used for determining
 the torques and resistance of airplane wings
 and was motivated by the developments of aeronautics
 during the First World War.

 \section{Acknowledgements}

We wish to thank Howard~Stone who made the conversation mentioned above possible
during the meeting of the Division of Fluid Dynamics of the American Physical Society
in Washington in November 2000.
We are also grateful to Michael~Brenner for bringing reference~\cite{DHMICHAEL}
and the beautiful papers~\cite{TORSION} to our attention.

\end{document}